\documentclass[12pt]{iopart}

\usepackage{graphicx}

\begin{document}
\title{An analytical model of  depth-dose distributions for carbon-ion beams
}

\author{Fulya Halıcılar$^{1,*} $ and Metin Arık$^{1}$}

\address{$^1$Department of Physics, Bogazici University, Bebek, Istanbul, Turkey – 34342
\\

\vspace{1.0 mm}
$^*$ Corresponding author E-mail: fulyahalicilar@gmail.com
\\
E-mail: metin.arik@boun.edu.tr

\vspace{1.0 mm}

}

\begin{abstract}

\textit{Objective.} Improving effective treatment plans in carbon ion therapy, especially for targeting radioresistant tumors located in deep-seated regions while sparing normal tissues, depends on a precise and computationally efficient dose calculation model. Although dose calculations are mostly performed using Monte Carlo simulations, the large amount of computational effort required for these simulations hinders their use in clinical practice. To address this gap, we propose, for the first time in the literature, an analytical model for the depth-dose distribution of carbon-ion beams by adapting and extending Bortfeld’s proton dose model.  \textit{Approach.} The Bortfeld model, widely used for proton therapy, was modified and expanded by introducing additional terms and parameters to account for the energy deposition and fragmentation effects characteristic of carbon ions. The proposed analytical model was implemented in MATLAB software to calculate depth-dose distributions of carbon-ion beams across the clinical therapeutic energy range of 100–430 MeV/u. In particular, the calculated results for carbon-ion energies of 280 MeV/u and 430 MeV/u were compared with Monte Carlo simulation results obtained using TOPAS to assess the precision of our model.  \textit{Main Results.} It is observed that the results of the proposed model are in good agreement with those of several analytical and experimental studies for clinical carbon-ion beams within the therapeutic energy range of 100–400 MeV/u.  At 280 MeV/u, the analytical model result exhibited strong consistency with the depth-dose curve produced by TOPAS Monte Carlo simulations. However, noticeable discrepancies appeared at higher energies, such as 430 MeV/u, particularly in the Bragg peak height and the dose fall-off. \textit{Significance.} In the clinically useful energy range, our analytical model could potentially be an effective tool in carbon ion therapy as an alternative to complex Monte Carlo simulations.  It will enable fast dose assessment with accuracy and in real-time, thus improving workflow efficiency. 

\end{abstract}

\maketitle

\section{Introduction}
Carbon-ion beams are a potent radiation therapy tool because of their distinctive depth-dose distribution characterized by the Bragg peak, which allows for precise targeting of the tumor with sparing of the surrounding healthy tissues (Durante and Loeffler 2010, Kamada \textit{et al} 2015, Durante \textit{et al} 2017). Most importantly, accelerated carbon ions demonstrate superior dose localization compared to conventional radiation modalities, leading to a reduced exposure of normal tissues located in front of the tumor and enhancing therapeutic efficiency (Kanai \textit{et al} 1997, Amaldi and Kraft 2005, Schulz-Ertner \textit{et al} 2006,  Mohamad \textit{et al} 2017). This benefit arises from the distinctive physical properties of carbon-ion beam, i.e., reduced range straggling and a sharper, narrower Bragg peak, with less dose deposition in the entrance region (Schardt \textit{et al} 2010, Newhauser and Zhang 2015). In addition, carbon-ion beams have lower lateral scattering and higher relative biological effectiveness (RBE) at the Bragg peak, which enables more effective destruction of the tumor cells with sparing of nearby normal structure (Tsujii and Kamada 2012, Durante and Paganetti 2016, Kim \textit{et al} 2020, Malouff \textit{et al} 2020, Tinganelli and Durante 2020). The enhanced dose localization and the higher linear energy transfer (LET) greatly increase the probability of tumor control, especially in the therapeutic management of radioresistant tumors (Kamada \textit{et al} 2015). These characteristics make carbon ion therapy a conformal and effective treatment modality with the potential to maximize tumor control and minimize normal tissue toxicity (Kamada \textit{et al} 2015, Mohamad \textit{et al} 2017, Kraft 2000, Tinganelli and Durante 2020).

These clinical effects depend on proper modeling of the complex nuclear interactions occurring when carbon ions travel through tissue. Of these interactions, peripheral collisions, in which carbon ions lose one or more nucleons due to fragmentation, are most common. These interactions are well described by the two-step abrasion-ablation model (Serber 1947, Hüfner \textit{et al} 1975, Gunzert-Marx \textit{et al} 2008, Schardt \textit{et al} 2010). In the abrasion phase, nucleons are stripped from both the target nuclei and incoming carbon ions in the overlapping zone, resulting in the formation of a thermally excited “fireball”.  In the subsequent ablation phase, the remnants undergo de-excitation and radiate light clusters and nucleons. Projectile fragments, with velocities equal to or larger than the entering beam, continue to deliver dose until they are brought to rest. The fragmentation process minimizes the initial beam flux and forms secondary fragments of lower atomic numbers. Since the range of the particle scales as A/Z², these fragments have longer ranges and may deliver undesired doses to healthy tissues beyond the Bragg peak. This effect, known as the carbon-ion fragmentation tail, is determined by carbon ion charge-changing cross sections in water (Schall \textit{et al} 1996, Golovchenko \textit{et al} 1999, Golovchenko \textit{et al} 2002). Lighter fragments, such as hydrogen and helium, create the tail after the Bragg peak (Haettner \textit{et al} 2013, Ying \textit{et al} 2017, Nandy 2021), while heavier fragments, such as carbon and boron, spread doses before and after the Bragg peak (Ying \textit{et al} 2019, Hamad 2021).

Accurate simulation of these fragmentation processes is important in optimal treatment planning of carbon ion radiotherapy, where dose distribution and biological effectiveness need to be controlled precisely. Monte Carlo simulation platforms like TOPAS (Perl \textit{et al} 2012, Liu 2017, Halıcılar \textit{et al} 2024), Geant4 (Agostinelli \textit{et al} 2003, Hamad 2021), FLUKA (Böhlen \textit{et al} 2014, Battistoni \textit{et al} 2015), and PHITS (Sato \textit{et al} 2018) are the essential resources for carbon-ion dose distribution predictions. Apart from simulating physical interactions and biological effects of ion beams, these platforms make it possible for full modeling of depth-dose profiles, fragmentation tails, LET spectra, and RBE distributions, providing useful information on enhancing physical as well as biological models used in carbon ion therapy (Robert \textit{et al} 2013, Battistoni \textit{et al} 2016). For example, TOPAS is applied in clinical studies since it is robust enough to simulate various situations of treatments (Faddegon \textit{et al} 2020), while Geant4 possesses superior capability in simulating nuclear interactions and secondary particle production (Allison \textit{et al} 2016). Similarly, PHITS (Niita \textit{et al} 2010, Furuta and Sato 2021) and FLUKA (Kozłowska \textit{et al} 2019) are known for their reliable performance in simulating ion beam transport in complex geometries and heterogeneous media. However, the considerable computational requirements and prolonged processing time of Monte Carlo simulations limit their clinical utility in routine procedures (Krämer \textit{et al} 2000).  This deficiency has given rise to analytical models that yield a less costly computational approach to dose calculation for effective treatment planning systems (TPS).

In this context, analytical methods have gained much interest since they can find a balance between computational efficiency and predictive accuracy. A prominent example is the Bortfeld Analytical Formulation (BAF) model, which was originally formulated to approximate the Bragg curve in proton therapy (Bortfeld 1997). The BAF model has been greatly developed and improved to make it more accurate and versatile in proton treatment applications. Zhang \textit{et al} (2011) established a parameterization method that describes multiple Bragg curves simultaneously to enhance the accuracy of models for scanning proton beams. Based on this, Zhang \textit{et al}(2022a) presented corrections to integral depth-dose calculations employing a straight scattering track approximation, while Zhang \textit{et al} (2023) proposed a secondary propagation model for dose contributions from the secondary particles. Kelleter and Jolly (2020) expressed the mathematical formulation of the depth-light curve model, considering quenching in scintillators and extending the scope of the model for optical dosimetry applications. Nichelatti \textit{et al} (2019) modified the BAF model for use with lithium fluoride to allow for precise energy distribution analysis in experimental arrangements. Similarly, Aminafshar \textit{et al} (2024a) extended the BAF model to biological tissues to make it more relevant to clinical treatment planning. Furthermore, Aminafshar \textit{et al} (2024b) advanced the model for heterogeneous media and provided realistic simulations of dose distributions for cases with anatomical complexity. These advancements highlight the BAF model’s increasing importance in developing analytical approaches for charged particle therapy.
\newpage
However, the conventional dose calculation models are still difficult to apply for carbon-ion beams because of the complicated physical and biological behavior of heavy ions with increased LET, pronounced nuclear fragmentation, and energy straggling, which make it difficult to accurately calculate doses (Krämer and Scholz 2000). Because of these complexities, various researchers have proposed analytical models specific to the carbon-ion beam behavior. Kempe and Brahme (2010) developed an analytical theory to calculate the fluence and absorbed dose from primary ions and their fragments in broad light-ion beams. Hollmark \textit{et al}(2008) presented a model that accounts for the multiple scattering of primary and secondary ions in light-ion pencil beams to enhance lateral dose predictions. Donahue \textit{et al} (2016) also provided a stopping power and range model for hydrogen and heavier ions in the therapeutic energy range. Kundrát (2007) developed a semi-analytical model that combines simplified dose calculations with a probabilistic approach to estimate biological effectiveness and enable more efficient, biologically informed treatment planning for light-ion radiotherapy. All these developments apart, an accurate but computationally efficient analytical model of carbon-ion depth-dose behavior over clinically useful ranges remained a significant challenge.

In response to these challenges, we present a novel analytical approach to calculate depth-dose distributions in carbon ion therapy, based on Bortfeld’s established proton dose calculation model. The proposed model incorporates carbon-ion properties, such as distinctive energy deposition modes and fragmentation effects, across the energy range of 100 to 430 MeV/u. Through systematic validation against TOPAS Monte Carlo simulations, the model showed high accuracy, especially at clinically meaningful energies of about 280 MeV/u. Although there are some discrepancies at 430 MeV/u, particularly in peak height and dose fall-off behavior, the proposed model's performance at conventional therapeutic energies makes it a useful tool for optimizing treatment planning. Our model has the potential to enhance carbon ion therapy treatment planning by balancing computational efficiency with dosimetric precision to enable more rapid generation of accurate depth-dose distributions for efficient tumor targeting. It is anticipated that our model will significantly enhance its clinical utility with further improvements and extensive validation studies in various clinical situations.

\section{Methods}
\subsection{Analytical derivation of a carbon-ion Bragg curve}

In this study, we introduce an analytical model for calculating the dose distribution of carbon-ion beams in a homogeneous medium with a focus on the Bragg curve. The proposed model is based on the Bortfeld model for proton beams (Bortfeld 1997), with additional terms to account for nuclear fragmentation processes characteristic of carbon-ion interactions.
\newpage
To establish the model, we assume that a monoenergetic carbon-ion beam is incident along the z-axis on a homogeneous medium at z = 0. The energy fluence at depth z within the medium is given by

\begin{equation}
{\Psi(z) = \Phi(z) E(z)},
	\label{eq: Psi}
\end{equation}
where $\mathrm{\ }\mathrm{\Phi }\left(z\right)$ represents the particle fluence and $E\left(z\right)$ denotes the residual energy at depth z.  The energy fluence 
$\Psi(z)$ is essential for determining the total energy imparted to the medium per unit mass.

In the Bortfeld model for proton beams, the absorbed dose equation in terms of the total energy released per unit mass is expressed as
\begin{equation}
	\setlength\abovedisplayskip{20pt} 
	\setlength\belowdisplayskip{20pt}
	\hat{D}\left(z\right)\mathrm{=-}\frac{\mathrm{1}}{\ \varrho }\left(\mathrm{\Phi }\left(z\right)\frac{dE(z)}{dz}+\gamma \frac{d\mathrm{\Phi }\left(z\right)}{dz}E\left(z\right)\right),
	\label{eq: Hat_dose}
\end{equation}
where $\varrho$ is the mass density of the medium and $\gamma$ is a certain fraction of locally absorbed energy released in non-elastic nuclear interactions.

The analytical model for a carbon Bragg curve can be obtained by modifying and adding new terms to the above formula. The total dose $\hat{D}$($z$) should contain the dose expression for the secondary fragments $\hat{D}_{sf}$($z$) in addition to the dose contribution of the primary ions $\hat{D}_{pp}$($z$). This yields
\begin{equation}
	\hat{D}\left(z\right)\mathrm{=}\hat{D}_{pp}\left(z\right)+{\hat{D}}_{sf}\left(z\right).
	\label{eq: Hatdose}
\end{equation}

\subsection*{Primary carbon-ions dose contribution}

The dose contribution from primary carbon ions arises from their energy deposition as they propagate through the medium. This energy loss is predominantly governed by the stopping power of the medium and elastic scattering. Due to their relatively high mass compared to protons, carbon ions transfer more energy to electrons during their passage. However, the secondary electrons produced have considerably shorter ranges compared to those of the primary carbon ions. Despite this, the energy deposited by both the primary carbon ions and the secondary electrons contributes substantially to dose deposition, especially in the region near the Bragg peak.

The dose of primary particles, as the total energy released per unit mass, can be given by
\begin{equation}
	\setlength\abovedisplayskip{20pt} 
	\setlength\belowdisplayskip{20pt}
	\hat{D}_{pp}\left(z\right)\mathrm{=-}\frac{\mathrm{1}}{\ \varrho }\left(\mathrm{\Phi_{pp} }\left(z\right)\frac{dE_{pp}(z)}{dz}+\gamma \frac{d\mathrm{\Phi_{pp} }\left(z\right)}{dz}E_{pp}\left(z\right)\right),
	\label{eq: dose_pp}
\end{equation}
where $\Phi_{pp}(z)$ is the fluence of the primary carbon ions and $E_{pp}(z)$ is the residual energy of the primary ions at depth 
z. The term ${dE_{pp}(z)}/{dz}$ represents the rate of energy loss due to the stopping power of the medium, while ${d\Phi_{pp}(z)}/{dz}$ accounts for the reduction in fluence, primarily due to elastic scattering. Elastic scattering causes multiple scattering, which affects the ion’s trajectory and broadens the beam, influencing the dose profile.

\subsection*{Secondary fragments dose contribution}

The secondary dose contribution arises from inelastic nuclear interactions, which include nuclear fragmentation processes. These interactions generate secondary fragments that contribute to the fragmentation tail observed both before and after the Bragg peak. The energy deposited by secondary fragments is critical for accurately predicting dose distribution, especially in the tail region.

The dose of secondary fragments is expressed as
\begin{equation}
	\setlength\abovedisplayskip{20pt} 
	\setlength\belowdisplayskip{20pt}
	\hat{D}_{sf}\left(z\right)\mathrm{=-}\frac{\mathrm{1}}{\ \varrho }\left(\mathrm{\Phi_{sf} }\left(z\right)\frac{dE_{sf}(z)}{dz}+\gamma \frac{d\mathrm{\Phi_{sf} }\left(z\right)}{dz}E_{sf}\left(z\right)\right),
	\label{eq: dose_sf}
\end{equation}
where $\Phi_{sf}(z)$ is the fluence of the secondary fragments, $E_{sf}(z)$ refers to the kinetic energy of the secondary fragments, which are produced from the nuclear fragmentation of the primary ions as they interact with the medium. The term ${dE_{sf}(z)}/{dz}$ refers to the rate of energy loss per unit depth experienced by the secondary fragments due to the medium's stopping power caused by ionization and energy transfer mechanisms.  The rate of change in fluence of secondary fragments, ${d\Phi_{sf}(z)}/{dz}$, is controlled by inelastic scattering caused by nuclear fragmentation.

In this study, the parameter $\gamma$, which denotes the fraction of energy locally absorbed within the medium, is set to 0.6 and is taken to be the same for both primary carbon ions and secondary fragments. Despite the fundamentally different physical mechanisms involved, this assumption makes it easier to apply the fluence-dependent correction term in the dose calculation consistently. Fluence reduction for primary carbon ions is specifically governed by elastic Coulomb scattering and nuclear interactions, which influence the trajectory and lateral spread of the ions.  In contrast, fluence reduction for secondary fragments results from inelastic nuclear fragmentation mechanisms, which lead to a reduction in the number of fragments reaching the target. The model streamlines the formulation, simplifies the complex interactions, and accounts for both elastic and inelastic effects by employing a unified $\gamma$  value for primary and secondary contributions. This approach enables a coherent treatment of the fluence-dependent dose contributions across different physical processes in addition to improving computational efficiency.

Inelastic scattering ends up in secondary fragments and contributes to the tail of the dose distribution beyond the Bragg peak. The dose contribution from secondary fragments can be modeled as the sum of the doses from various secondary fragments generated through nuclear fragmentation. The contribution of dose for secondaries can be described as
\begin{equation}
    \hat{D}_{sf}(z) =
    \cases{
        \hat{D}_{sf}^{h}(z) & \textrm{for $z \leq R_0$}, \\
        \hat{D}_{sf}^{l}(z) + \hat{D}_{sf}^{h}(z) & \textrm{for $z \geq R_0$},
    }
    \label{eq: dose_s}
\end{equation}
where $\hat{D}_{sf}^{l}(z)$ and $\hat{D}_{sf}^{h}(z)$ represent the contributions from lighter and heavier fragments, respectively, $R_0$
 denotes the range of primary carbon ions.
\newpage
Secondary fragments, such as boron (B), beryllium (Be), hydrogen (H), helium (He), and lithium (Li), are produced during the fragmentation of carbon ions in the target. These secondary particles have broader energy spectra and longer ranges compared to the primary carbon ions. This is due to the scaling of the particles’ range with ${A}/{{{Z}}^2}$ at the same velocity (Ying \textit{et al} 2017, Nandy 2021), and these secondary fragments contribute to the tail dose beyond the Bragg peak. A notable example is carbon-11 ($^{11}\mathrm{C}$), which originates both from the fragmentation of incident $^{12}\mathrm{C}$ ions and from interactions with target nuclei, thereby contributing to the dose distribution before the Bragg peak (Gunzert-Marx \textit{et al} 2008).

In this study, we explicitly model the dose contributions from heavier and lighter secondary fragments. The dose contribution from heavier secondary fragments is expressed as 

\begin{equation}
    \hat{D}_{sf}^{h}(z) =
    \cases{
        \hat{D}_{^{11}\mathrm{B}}(z) + \hat{D}_{^{11}\mathrm{C}}(z) & \textrm{for $z \leq R_0$} , \\
        \hat{D}_{^{10}\mathrm{B}}(z) & \textrm{for $z \geq R_0$ .}
    }
    \label{eq: dsfhf1}
\end{equation}
Similarly, the dose contribution from lighter secondary fragments is given by
\begin{equation}
\hat{D}_{sf}^{l}\left(z\right)\mathrm{=}\hat{D}_{\mathrm{p}}\left(z\right)+\hat{D}_{\mathrm{He}}\left(z\right).
	\label{eq:dsflf1}
\end{equation}
Here, the terms $\hat{D}_{^{11}\mathrm{B}}(z)$ and $\hat{D}_{^{11}\mathrm{C}}(z)$ refer to the dose contributions from  $^{11}\mathrm{B}$ (Ying \textit{et al} 2019) and  $^{11}\mathrm{C}$ (Pschenichnov \textit{et al} 2010) fragments, respectively. These fragments are mainly produced by nuclear fragmentation of primary carbon ions and contribute to the dose distribution before the Bragg peak. Beyond the Bragg peak, $\hat{D}_{^{10}\mathrm{B}}(z)$, the dose contribution from  $^{10}\mathrm{B}$ (Francis \textit{et al} 2014) fragments becomes the dominant contributor to the dose profile due to its significant role as a fragmentation product of carbon ions. $\hat{D}_{\mathrm{p}}(z)$ and $\hat{D}_{\mathrm{He}}(z)$ reflect the dose contributions from lighter nuclei produced in nuclear fragmentation, i.e., protons (Gunzert-Marx \textit{et al} 2008) and helium nuclei (He) (Haettner \textit{et al} 2013). Because of their long enough ranges, protons and helium nuclei can both have a considerable impact on the dose profile beyond the Bragg peak, which can affect the tail region. Consequently, the total dose distribution is shaped not only by the primary carbon ions but also by the secondary fragmentation products, including boron-10, protons, and helium nuclei, each contributing at different depths and ranges within the irradiated tissue.

\subsection*{Stopping power and range calculations for carbon ions}

The relationship between the initial energy of the carbon ion $E\left(z\mathrm{=0}\right)=$ $E_{0\ }$and the range $z\mathrm{=}R_0$ is determined by the Bragg-Kleemann rule (Bragg and Kleemann 1905, Evans 1985). In heavy ions, this rule is obtained by multiplying with the factor  ${A}/{{{Z}}^2}$ (Schardt \textit{et al} 2010). The range of the primary ion at depth z is therefore given by 
\begin{equation}
	{\ \ R}_0\mathrm{=}\frac{A}{\ {{Z}}^2}\ \alpha \ {E_{0\ }}^p,
	\label{eq: R_0}
\end{equation}
where $A$ is the mass number,  ${Z}$ is the atomic number, $\alpha $ and $p$ are the fitting parameters.

The Bragg-Kleemann rule provides a relationship between the energy of a charged particle and its range in a given medium. The penetration depth $R_0$ of heavy ions is determined by both the ion's mass number A and its atomic number ${Z}$. We provide ${A}/{{{Z}}^2}$ to account for the fluctuations in ion mass and charge. The values of $\alpha$ and p are determined from experimental data and appropriately describe the non-linear connection between energy and range as shown in Figure 1. Specifically, 1.76 is observed for the value of p for carbon ions. The coefficient 0.0007 is the product of the fitting parameter $\alpha$ and the factor ${A}/{{{Z}}^2}$. This equation allows one to predict the amount of ions that will penetrate tissue or other media for applications such as ion beam therapy in cancer treatment.
\begin{figure} [htpt]
	\centering
	\includegraphics[width=0.73\columnwidth]{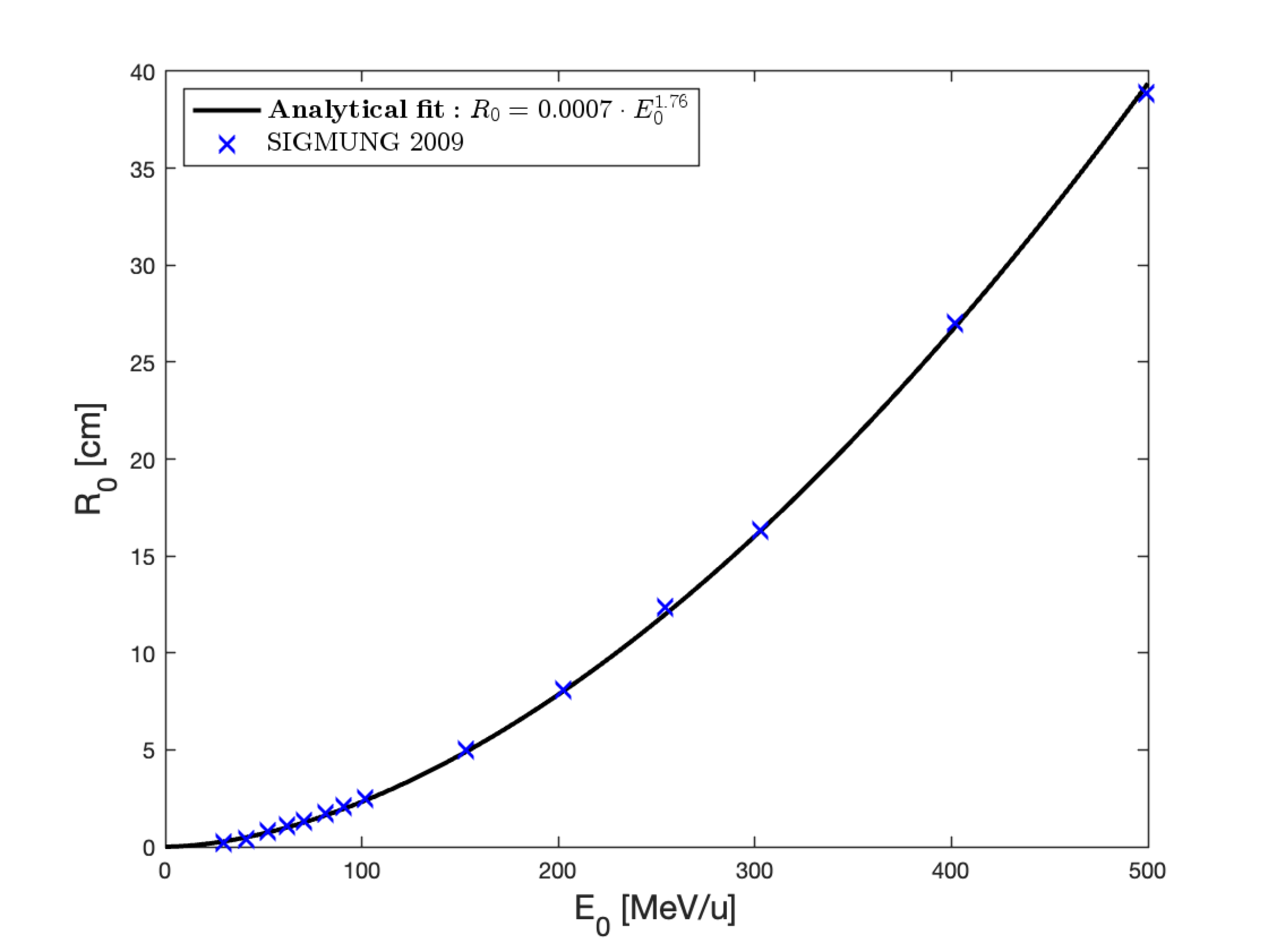}
	\caption{Comparing carbon-ion range data (up to 500 MeV) in water as a function of initial kinetic energy between SIGMUND 2009 and the proposed analytical model for carbon-ion beam. }
	\label{fig: Range_energy}
\end{figure}

A Gaussian distribution defines the range of the radiation-matter interaction. Carbon-ion beams can penetrate as deep as 40 cm and reach energies as high as 500 MeV/u, as Figure 1 illustrates. This formulation is consistent with empirical findings (Sigmund \textit{et al} 2009), as seen by the penetration depth of carbon ions in Figure 1, where deeper penetrations are correlated with higher energies, thereby validating the theoretical model.

As a carbon ion traverses the medium, it loses energy due to ionization and other radiation-matter interactions. This continuous energy loss persists until the ion reaches the end of its range, where its energy is nearly depleted, and the Bragg peak is observed. The residual energy of the primary carbon ions, $E_{pp}(z)$, decreases progressively as the ions penetrate the tissue at any depth z. This reduction in energy is a direct result of ionization and other interactions along the ion’s trajectory.

The carbon-ion beam deposits energy along its path, extending from the surface ($z=0$) to the maximum range ($z=R_0$). At any given depth z within the medium, where \textit{z}$\ \leq R_0\ $, the remaining energy required for the primary ions to travel the distance $R_0-z$ can be expressed by the rearranged range-energy relationship. Specifically, the residual energy of the primary carbon ions at depth z is given by

\begin{equation}
	\setlength\abovedisplayskip{20pt} 
	\setlength\belowdisplayskip{20pt}
	\ E_{pp}\left(z\right)=\frac{1}{\alpha^{1/p}}{\left(\frac{A}{\ {{Z}}^2}\right)}^{{-1}/{p}}{\left(R_0\ -\ z\right)}^{{1}/{p}}.
	\label{eq: energy_pp}
\end{equation}

The stopping power, which describes the rate at which the ion loses energy per unit distance traveled, is related to the residual energy and is highest near the Bragg peak, where the ion loses energy most rapidly. The stopping power for primary carbon ions, thus, can be written as
\begin{equation}
	\setlength\abovedisplayskip{20pt} 
	\setlength\belowdisplayskip{20pt}
	S_{pp}\left(z\right)=-\frac{dE_{pp}}{dz}=\frac{1}{p\,\alpha^{1/p}}{\left(\frac{A}{\ {{Z}}^2}\right)}^{{-1}/{p}}{(R_0\ -\ z)}^{{1}/{p-1}} .
	\label{eq: stopping_pp}
\end{equation}

In addition to the primary carbon ions, secondary fragments from nuclear interactions between the carbon ion and the medium are also generated. Secondary fragments such as protons, neutrons, or light nuclei possess unique energy deposition properties that can be approximated in terms of concepts similar to those of the primary carbon ions. However, the secondary fragment kinetic energy and stopping power are adjusted according to the fragment's specific mass (A) and charge (Z). The kinetic energy of any secondary fragment at any depth in the medium is determined by an equation similar to that of the original carbon ion but with adjustments according to the fragment charge and mass. As the fragment moves through the medium, its kinetic energy loss with depth varies as a function of its particular nuclear characteristics. The process is described as a function of the range of the fragment, adjusted for its mass and charge compared to the primary ion.

Likewise, the stopping power of the secondary fragments, which determines the rate at which energy is lost per unit distance, is of the same form as that for the primary carbon ions. The stopping power for secondary fragments is larger due to the fragments' lower mass and hence loss per unit distance traveled. Therefore, the secondary fragments have a greater rate of energy deposition when passing through the medium and contribute largely to the dose profile contained in the fragmentation tail beyond the Bragg peak.

In summary, although the stopping power and secondary fragment kinetic energy have the same rules as primary carbon ions, their A and Z values numerically provide different patterns of energy loss. These differences are of major importance for the determination of the total dose distribution in carbon ion therapy,  since both primary ions and secondary fragments contribute to total dose deposition. Secondary fragments significantly influence the formation of the fragmentation tail observed after the Bragg peak, making the dose profile more complicated.

\subsection*{Impact of energy loss and nuclear fragmentation on beam fluence}

During passage of carbon-ion beams through biological tissue or other materials, the primary ions lose energy continuously by electromagnetic interactions and nuclear reactions (Schardt et al 2010, Sigmund 2014). Electronic stopping is the major mechanism of energy dissipation, where ionization and excitation of atomic electrons result in a gradual reduction in beam energy (Ziegler \textit{et al} 2010). Yet, nuclear interactions, i.e., inelastic collisions with the target nuclei, introduce additional complexity by causing fragmentation of the primary ions (Schardt \textit{et al} 2010, Sigmund 2014). These nuclear interactions are responsible for a redistribution of fluence between the secondary particles, leading to an attenuation of the primary beam (Haettner \textit{et al} 2013). Nuclear fragmentation creates secondary fragments with different charge to mass ratios and can continue traveling in the medium with changed trajectory and energy spectra (Matsufuji et al 2003). Consequently, fluence attenuation in carbon-ion therapy is influenced not only by energy loss but also by the gradual depletion of the primary particle population due to nuclear fragmentation (Schall \textit{et al} 1996).

Nuclear interactions are the major reason for fluence decrease along the beam path. When the carbon-ion beam is near the end of its range in the medium, nuclear reaction is more likely to take place, resulting in a marked decrease in primary particle fluence. This process can be mathematically described by considering the attenuation of the primary particle fluence $\Phi_{pp}$ as a function of depth z, given by

\begin{equation}
	{\Phi _{pp}}={\Phi_0}\;exp({-\mu^p}z),
	\label{eq:Phipp}
\end{equation}
where $\Phi_{0}$ represents the initial fluence at the entrance of the medium, $p$  and $ \mu^p = \beta$ are empirical parameters introduced by Bortfeld to characterize fluence attenuation. This expression accounts for the exponential reduction in primary fluence as the beam propagates through the medium.

To gain a better insight into the effect of nuclear fragmentation on fluence attenuation, the particle fluence can be approximated using the first two terms of a Taylor series expansion:

\begin{equation}
	{\Phi_{pp} }\left(z\right)\propto 1+\ \beta  {(R}_0\ -\ z),
	\label{eq: phi}
\end{equation}

where $R_0$ is the initial range of the carbon-ion beam. Normalizing this expression for primary particle fluence $\Phi_{0}$ yields

\begin{equation}
	\setlength\abovedisplayskip{10pt} 
	\setlength\belowdisplayskip{10pt}
	{\Phi_{pp} }\left(z\right)={\Phi }_{0 }\frac{1+\beta ({(R}_0\ -\ z)}{1+\beta R_0} .
	\label{eq: Phi_pp}
\end{equation}

This expression provides a practical approximation of fluence reduction, particularly in the near Bragg peak region, where nuclear fragmentation effects become more pronounced.

 To accurately model depth-dose deposition for carbon ions, the rate of fluence reduction must also be considered. The differential expression governing the variation in fluence with depth is given by
\begin{equation}
	\setlength\abovedisplayskip{10pt} 
	\setlength\belowdisplayskip{10pt}
	-\ \frac{d{\Phi_{pp} }}{dz}={\Phi }_{0}\frac{\beta }{1+\beta R_0}. 
	\label{eq: dPhi}
\end{equation}
This equation quantifies the attenuation of primary particle fluence as a function of depth, incorporating both energy loss mechanisms and the depletion of primary ions due to nuclear fragmentation.

In particular, the mathematical framework used to define the fluence and fluence reduction of primary particles can be directly applied to the fluence and fluence reduction of secondary fragments, considering each fragment type. This approach is needed to forecast the fluence reduction of secondary fragments, which is a crucial function of the total depth-dose distribution of carbon ions. In contrast to proton beams, where nuclear interactions are less prominent, carbon-ion beams exhibit a complex fragmentation tail beyond the Bragg peak. This tail arises from the residual energy carried by the secondary fragments, which continue to deposit dose at greater depths compared to the primary ions. Understanding and accurately modeling these effects are essential for optimizing treatment planning in carbon-ion therapy.

\subsection*{Depth-dose distribution in absence of range straggling}

The depth-dose distribution can be obtained by incorporating the residual energy and stopping power of the primary particles, the kinetic energy and stopping power of the secondary fragments, as well as the fluence and fluence reduction of both the primary particles and secondary fragments into Equation (3) for the total dose. This can be expressed as

\begin{equation}
    \hat{D}(z) =
    \cases{
        \hat{D}_{pp}(z) + \hat{D}_{sf}^{h}(z) & \textrm{for $z \leq R_0$}, \\
        \hat{D}_{sf}^{l}(z) + \hat{D}_{sf}^{h}(z) & \textrm{for $z \geq R_0$},
    }
    \label{eq:hatdose_pp}
\end{equation}

where the dose contributions are defined as:

\begin{equation}
    \hat{D}_{sf}^{h}(z) =
    \cases{
        \hat{D}_{^{11}\mathrm{B}}(z) + \hat{D}_{^{11}\mathrm{C}}(z) & \textrm{for $z \leq R_0$} , \\
        \hat{D}_{^{10}\mathrm{B}}(z) & \textrm{for $z \geq R_0$ ,}
    }
    \label{eq: dsfhf}   
\end{equation}

\begin{equation}
\hat{D}_{sf}^{l}\left(z\right)\mathrm{=}\hat{D}_{\mathrm{p}}\left(z\right)+\hat{D}_{\mathrm{He}}\left(z\right),
	\label{eq:dsflf}
\end{equation}
  
\begin{eqnarray}
\hat{D}_{pp}(z) &=& \frac{\Phi_0}{\varrho \left(\frac{A_{^{12}\mathrm{C}}}{Z_{^{12}\mathrm{C}}^2} \alpha\right)^{1/p} (1 + \beta R_0)} \nonumber\\
    && \times \left[ \frac{1}{p} (R_0 - z)^{\frac{1}{p} - 1} + \left( \frac{\beta}{p} + \gamma \beta  \right) (R_0 - z)^{\frac{1}{p}} \right].
    \label{eq:hatDpp}
\end{eqnarray}

$\hat{D}_{pp}(z)$ accounts for the dose delivered by primary carbon ions that have undergone elastic nuclear interactions, and it follows the stopping power characteristics of the carbon ions. This dose results in a sharp Bragg peak at $R_0$, which is a typical feature of ion beams, although range straggling is neglected in this simplified approach.

The dose from the secondary fragments, $\hat{D}_{sf}(z)$ arises from inelastic nuclear interactions occurring at various depths within the water. The contributions from heavier secondary fragments, such as boron and carbon isotopes, dominate the dose at depths below the Bragg peak ($z \leq R_0$), while lighter secondary fragments, such as protons and helium nuclei, extend the dose distribution beyond the Bragg peak ($z \geq R_0$).

The dose contribution from the secondary fragments $\hat{D}_{sf}(z)$  can be described by the following expression:
\begin{eqnarray}
    \hat{D}_{sf}(z) &=& \frac{\Phi_0}{\varrho \left(\frac{A_{sf}}{Z_{sf}^2}\alpha\right)^{1/p} (1 + \beta R_0)} \nonumber\\
    && \times \left[ \frac{1}{p} (R_0 - z)^{\frac{1}{p} - 1} + \left( \frac{\beta}{p} + \gamma \beta  \right) (R_0 - z)^{\frac{1}{p}} \right].
    \label{eq:hatDsf}
\end{eqnarray}

 If the density of the medium is given in ${\mathrm{g}}/{{\mathrm{cm}}^3}$, the $\hat{D}$ unit of the absorbed dose is MeV/$\mathrm{g}$. To write the unit of the dose as Gray (energy per unit mass), we need to multiply the expression by the factor ${\mathrm{10}}^9$ \textit{e}/\textit{C}=1.602${\mathrm{\times 10}}^{-10}$.

\subsection*{Impact of range straggling on the depth-dose curve of carbon-ion beams}

In the process of energy loss during the slowing down of carbon-ion beams in water, statistical fluctuations occur in the energy loss of the ions due to multiple collisions. This phenomenon, known as energy-loss straggling, leads to an expansion of the carbon-ion beam. These fluctuations are governed by the Vavilov distribution (Vavilov 1957), which characterizes the probability distribution of energy loss in many collisions. In certain limits, particularly for high-energy beams, the Vavilov distribution approximates a Gaussian function (Bethe and Ashkin 1953, Bohr 1940, Ahlen 1980).

Since carbon ions with the same initial energy $E_{0}$ experience variations in energy loss, their stopping points differ, leading to range straggling. This effect follows a Gaussian distribution with a standard deviation ${\sigma }_z(\overline{z}$)=$\sigma $ centered around the mean depth $\overline{z}$. As a result, the dose formula as the convolution integral with the Gaussian distribution can be written in the form
\begin{equation}
	\setlength\abovedisplayskip{20pt} 
	\setlength\belowdisplayskip{20pt}
	{D}_{pp}(z) = \left\langle \hat{D}_{pp} \right\rangle (z) = \frac{1}{\sqrt{2\pi} \sigma} \int_{-\infty}^{R_0} \hat{D}_{pp}(\overline{z}) e^{-\frac{(z - \overline{z})^2}{2\sigma^2}}d\overline{z}.
	\label{eq: dose_gaussian}
\end{equation}
The partial integration method is obtained by simulating the integrals (Gradshteyn and Ryzhik  1980). The resulting dose, $\ {D}_{pp}\left(z\right)$, is given by
\begin{eqnarray}
    {D}_{pp}(z) &=& \frac{\Phi_{0} \ e^{-\zeta^2/4} \sigma^{1/p} \Gamma(1/p)}{\sqrt{2\pi} \varrho \left(\frac{A_{^{12}\mathrm{C}}} {Z_{^{12}\mathrm{C}}^2} \alpha\right)^{1/p} (1 + \beta R_0)} \nonumber\\
    && \times \left[ \frac{1}{\sigma} \mathcal{D}_{-\frac{1}{p}}(-\zeta) + \left( \frac{\beta}{p} + \gamma \beta \right) \mathcal{D}_{-\frac{1}{p}-1}(-\zeta) \right],
    \label{eq: Absorbed_Dose_pp}
\end{eqnarray}

where $\mathrm{\Gamma }$($\chi $) is the gamma function, ${\mathcal{D}}_y$($\chi $) is parabolic cylinder function (Abramowitz and Stegun 1972), ${\mathrm{\Phi }}_{\mathrm{0\ \ }}\mathrm{is\ the}$ initial fluence, $\sigma$ is the Gauss width of range diffusion and $\zeta ={(R_0-z)}/{\sigma }$.

For initial mono-energetic carbon beams, the standard deviation of range straggling, $\sigma$ = ${\sigma }_{mono}$, depends on the initial energy, $E_{0\ }$, or range $R_0$. The standard variance of range straggling, as derived by Chu (Chu \textit{et al} 1993), is given by
\begin{equation}
	{\sigma }_{mono}= 0.012 {R_0}^{0.951}A^{-0.5}.
	\label{eq: sigma_mono}
\end{equation}

This expression is in good agreement with the analytical model developed by Hollmark \textit{et al} (2004). For example, when the radiation of the carbon ion with energy 391MeV/u was considered, its range was found to be $R_0$ = 26.86 cm, and the standard deviation was calculated to be ${\sigma }_{mono}$ = 0.08 cm (Chu \textit{et al} 1993).

It is important to emphasize that the secondary fragments produced during nuclear interactions of carbon ions with tissue do not experience range straggling in the same manner as the primary carbon ions. Range straggling is mainly associated with the primary carbon ions, which undergo multiple collisions and energy losses as they travel through the medium. However, secondary fragments, such as protons, helium nuclei, or heavier fragments like boron and carbon isotopes, are produced at various depths in the tissue as a result of nuclear interactions. These secondary fragments generally have a higher velocity and much lower energy compared to the primary carbon ions. As a result, they do not undergo significant energy-loss fluctuations or range straggling in the same way the primary ions do. Their ranges are more localized and deterministic, with minimal spread around the mean value.
Therefore, secondary fragments do not experience the same range broadening due to energy-loss straggling. Their range remains consistent and concentrated, and they do not contribute to the Gaussian convolution typically employed to model range straggling in dose distribution calculations. 

\subsection*{Energy spectrum analysis}
In previous analyses, the carbon ion beam has been idealized as mono-energetic. However, this assumption does not accurately reflect real experimental conditions. In practice, carbon-ion beams exhibit an inherent spectral energy distribution, which is influenced by the characteristics of the accelerator, beam transport system, and collimation setup. A natural approach to account for the energy spread is to model the spectrum as a weighted superposition of mono-energetic Bragg curves with corresponding weight factors $\Phi_E$. In general, no exact analytical solution exists for this scenario, necessitating the use of numerical methods. In this work, we adopt an approximation that permits an analytical treatment of the problem, leading to a solution of the form given by Eq.(22).

Typical energy spectra of carbon ion beams consist of a prominent peak that can be approximated by a Gaussian energy distribution centered at  ${E=E}_{0\ }$, along with a relatively small tail extending towards both the peak and the lower energies. While the low-energy tail contributes to deviations from an ideal Gaussian shape, for analytical purposes, the distribution is often characterized by the energy straggling width, denoted as  ${\sigma }_{E,0}$, which is the standard deviation of the energy spectrum.

The range-energy relationship can be linearized around ${E=E}_{0\ }$due to the fact that ~${\sigma }_{E,0}\ll E_{0\ }$. Thus, the total straggling width, incorporating both energy and range components, can be obtained by summing the variances of the energy straggling width ${\sigma }_{E}\ $ and the range straggling width ${\sigma }_{mono}\ $. This yields
\begin{equation}
	\setlength\abovedisplayskip{15pt} 
	\setlength\belowdisplayskip{15pt}
	{\sigma }^2={\sigma }^2_{mono}+{\sigma }^2_{E,0}{\left(\frac{dR_0}{dE_{0\ }}\right)}^2={\sigma }^2_{mono}+{\ \left(\frac{A_{^{12}\mathrm{C}}}{Z_{^{12}\mathrm{C}}^2}\alpha \right)}^2p^2E_{0}^{2p-2}. 
	\label{eq: sigma2}
\end{equation}
It is complicated to calculate the tail of the energy spectrum due to many factors, and the shape cannot be precisely known. Therefore, if the total fluence corresponding to the tail is expressed by $\epsilon$, we can determine the spectrum of the tail by approximately simple modeling. ${\mathrm{\Phi }}_E\left(E\right)\ $is negligible at $E$=0 and a simple model with a positive slope for small $E$, i.e. ${\mathrm{\Phi }}_E(E)\mathrm{\propto }$ $E\ \mathrm{for\ }0\ll E\ll E_{0\ }$, is possible. By normalizing the integral of ${\mathrm{\Phi }}_E(E)$, we obtain
\begin{equation}
	{\Phi }_E\left(E\right)\mathrm{=}\epsilon {\mathrm{\Phi }}_0\frac{2E}{E_{0}^2} .
	\label{eq:phi_E}
\end{equation}
To calculate the depth dose distribution based on the linear energy spectrum, we have to translate ${\mathrm{\Phi }}_E\mathrm{(}E\mathrm{)}$ to ${\mathrm{\Phi }}_R\mathrm{(}R\mathrm{)}$. By using the range-energy relationship, ${\mathrm{\Phi }}_R\left(R\right)$ is given by

\begin{equation}
	\setlength\abovedisplayskip{10pt} 
	\setlength\belowdisplayskip{10pt}
	{\Phi }_R\left(R\right)={\mathrm{\Phi }}_E\left(E(R)\right)\frac{dE}{dR}=\epsilon {\mathrm{\Phi }}_0\frac{2{\
    Z_{^{12}\mathrm{C}}^{4/p}\ R}^{{2}/{p\ }-1}}{{\ \left(\frac{R_0 \ Z_{^{12}\mathrm{C}}^{2}}
{\alpha {{A_{^{12}\mathrm{C}}}}} \right)}^{{2}/{p}}p\ \alpha^{{2}/{p}}A_{^{12}\mathrm{C}}^{{2}/{p}}}.
	\label{eq: phi_R}
\end{equation}

For this expression, the range spectrum derived from the linear energy spectrum is approximately constant ($\left|{2}/{p\ }-1\right|\ll 1$). After assuming \textit{p}$\approx 2$\textit{ }we obtain
\begin{equation}
	{\Phi }_R\left(R\right)\approx constant=\epsilon {\mathrm{\Phi }}_0\frac{1}{R_0}.
	\label{eq: phi_Rapp}
\end{equation}
For carbon ion beams, the total absorbed depth-dose distribution can be derived by considering contributions from both primary carbon ions and secondary nuclear fragments. Firstly, we determine the contribution of primary carbon ions that have undergone elastic nuclear interactions. The depth dose of the tail of the primary carbon ions energy spectrum,$\ {\hat{D}_{pp}^{tail}},\ $can thus be approximated by
\begin{equation}
	 {\hat{D}_{pp}^{tail}}(z)\approx \frac{1}{{\mathrm{\Phi }}_0}\int^{R_0}_z{{\mathrm{\Phi }}_R\left(R\right)\ \hat{D}_{pp}\left(z,R\right)dR\ }.
	\label{eq: taildose}
\end{equation}
 The depth dose distribution for the tail is computed using a convolution integral similar to Eq. (28). Assuming that the dose from these ions is given as a function of z and R, we have
 \newpage
\begin{eqnarray}
    {\hat{D}_{pp}^{tail}}(z) &\approx& \frac{\epsilon \Phi_{0}}{R_0 \varrho p \left(\frac{A_{^{12}\mathrm{C}}}{Z_{^{12}\mathrm{C}}^2} \alpha \right)^{\frac{1}{p}}(1+\beta R_0)} 
    \int^{R_0}_z (R-z)^{\frac{1}{p}-1} dR \nonumber \\
    && + \frac{\epsilon \Phi_{0} (\beta + \gamma \beta p)}{R_0 \varrho p \left(\frac{A_{^{12}\mathrm{C}}}{Z_{^{12}\mathrm{C}}^2} \alpha \right)^{\frac{1}{p}}(1+\beta R_0)} 
    \int^{R_0}_z (R-z)^{\frac{1}{p}} dR, \\
    &=& \frac{\epsilon \Phi_{0}}{R_0 \varrho p \left(\frac{A_{^{12}\mathrm{C}}}{Z_{^{12}\mathrm{C}}^2} \alpha \right)^{\frac{1}{p}}(1+\beta R_0)} 
    (R_0-z)^{\frac{1}{p}}  \nonumber \\
    && + \frac{\epsilon \Phi_{0} (\beta + \gamma \beta p)}{R_0 \varrho p \left(\frac{A_{^{12}\mathrm{C}}}{Z_{^{12}\mathrm{C}}^2} \alpha \right)^{\frac{1}{p}}(1+\beta R_0)(1+p)}
     (R_0-z)^{\frac{1}{p}+1} .
    \label{eq: dose_app}
\end{eqnarray}

 We obtain the resulting dose calculation of carbon ions by adding the final contribution $ {\hat{D}_{pp}^{tail}}(z)$ to the absorbed dose $ {\hat{D}_{pp}}$\textit{(z}) of Eq.(19):
\begin{eqnarray}
    \hat{D}_{pp}(z) &=& \frac{\Phi_0}{\varrho \left(\frac{A_{^{12}\mathrm{C}}}{Z_{^{12}\mathrm{C}}^2} \alpha \right)^{\frac{1}{p}} (1 + \beta R_0)} \Bigg[ \frac{1}{p} (R_0 - z)^{\frac{1}{p} - 1} \nonumber \\
    && + \left( \frac{\beta}{p} + \gamma \beta + \frac{\epsilon}{R_0} \right) (R_0 - z)^{\frac{1}{p}} \nonumber \\
    && + \frac{\epsilon (\beta + \gamma \beta p)}{R_0 (1 + p)} (R_0 - z)^{\frac{1}{p} + 1} \Bigg].
    \label{eq:Dpphat}
\end{eqnarray}

Straggling can now be incorporated in a manner analogous to the methodology presented in the previous section, and the absorbed dose ${{D}_{pp}}$\textit{(z}) is thus found by the following expression:

\begin{eqnarray}
    D_{pp}(z) &=& \frac{\Phi_{0} \ e^{-\zeta^2/4} \sigma^{1/p} \Gamma(1/p)}{\sqrt{2\pi} \varrho \left(\frac{A_{^{12}\mathrm{C}}}{Z_{^{12}\mathrm{C}}^2} \alpha \right)^{\frac{1}{p}}(1 + \beta R_0)} \nonumber\\
     && \times \Bigg[ \frac{1}{\sigma} \mathcal{D}_{-\frac{1}{p}}(-\zeta) + \left( \frac{\epsilon}{R_0} + \frac{\beta}{p} + \gamma \beta \right) \mathcal{D}_{-\frac{1}{p}-1}(-\zeta) \nonumber \\
    && + \frac{\epsilon \sigma}{R_0} \left( \frac{\beta}{p} + \gamma \beta \right) \mathcal{D}_{-\frac{1}{p}-2}(-\zeta) \Bigg].
    \label{eq:dose_pp}
\end{eqnarray}
On the other hand, the secondary fragments, which arise from nuclear interactions between carbon ions and tissue nuclei, contribute a tail dose that extends beyond the Bragg peak. However, due to their lower mass and charge, these fragments do not experience significant range straggling. Their dose contribution can be approximated by
\newpage
\begin{eqnarray}
    \hat{D}_{sf}(z) &=& \frac{\Phi_0}{\varrho \left(\frac{A_{sf}}{Z_{sf}^2} \alpha \right)^{\frac{1}{p}} (1 + \beta R_0)} \Bigg[ \frac{1}{p} (R_0 - z)^{\frac{1}{p} - 1} \nonumber \\
    && + \left( \frac{\beta}{p} + \gamma \beta + \frac{\epsilon}{R_0} \right) (R_0 - z)^{\frac{1}{p}} \nonumber \\
    && + \frac{\epsilon (\beta + \gamma \beta {p})}{R_0 (1 + {p})} (R_0 - z)^{\frac{1}{p} + 1} \Bigg].
    \label{eq:hatdose_C11}
\end{eqnarray}
Unlike primary ions, secondary fragments do not experience range straggling because they are produced with a broad energy spectrum and continue their propagation with minimal additional energy fluctuations. These fragments contribute to an extended tail in the depth-dose distribution, which is particularly relevant for dose calculations in normal tissues beyond the tumor region.

To construct a complete analytical approximation of the depth-dose profile for carbon-ion beams, the total absorbed dose ${D}(z)$ is expressed as a piecewise function:

\begin{equation}
    D(z) \approx
    \left\{
    \begin{array}{ll}
        \hat{D}_{pp}(z) + \hat{D}_{sf}^{h}(z) & \textrm{for $  z\leq R_0 - 10\sigma$}, \\
        D_{pp}(z) & \textrm{for $R_0 - 10\sigma \leq z \leq R_0 + 5\sigma$}, \\
        \hat{D}_{sf}^{l}(z) + \hat{D}_{sf}^{h}(z) & \textrm{for $z \geq R_0 + 5\sigma$}.
    \end{array}
    \right.
    \label{eq: Dose_z}
\end{equation}

Here, $\hat{D}_{sf}^{l}(z)$ and $\hat{D}_{sf}^{h}(z)$ represent the contributions from lighter and heavier fragments, respectively:

\begin{equation}
    \hat{D}_{sf}^{h}(z) =
    \cases{
        \hat{D}_{^{11}\mathrm{B}}(z) + \hat{D}_{^{11}\mathrm{C}}(z) & \textrm{for $  z\leq R_0 - 10\sigma$}, \\
        \hat{D}_{^{10}\mathrm{B}}(z) & \textrm{for $z \geq R_0 + 5\sigma$},
    }
    \label{eq: dsfh}
\end{equation}
\begin{equation}
\hat{D}_{sf}^{l}\left(z\right)\mathrm{=}\hat{D}_{\mathrm{p}}\left(z\right)+\hat{D}_{\mathrm{He}}\left(z\right).
	\label{eq:dsfl}
\end{equation}

This formulation provides an analytical approximation for the depth-dose distribution of carbon-ion therapy, accurately modeling the contributions of both primary ions and secondary nuclear fragments. The methodology ensures consistency with physical interactions governing energy loss and nuclear fragmentation processes in heavy-ion therapy.

\subsection{TOPAS simulation}

Monte Carlo (MC) simulations of the Geant4 toolkit are widely used for calculating carbon-ion beam depth-dose distributions in water because they accurately model electromagnetic and nuclear interactions with high precision. Among these, the TOPAS  platform, built on Geant4, provides a versatile and efficient interface for configuring therapeutic beamlines and patient-like geometries (Perl \textit{et al} 2012). Several studies have used TOPAS in the calculation of depth-dose distributions of carbon-ion beams in water, ranging from energies of clinical interest.  For example, Liu \textit{et al} (2017) used TOPAS to perform the first Monte Carlo simulation of a carbon ion radiotherapy facility treatment head for beam energies between 100 and 400 MeV/u, whereas Yoon \textit{et al} (2023) integrated a modified microdosimetric kinetic model into the matRad treatment planning system and validated it against Geant4/TOPAS Monte Carlo simulations for carbon-ion beams in the range of 100–430 MeV/u. These experiments verify the reliability of Geant4-based MC algorithms, especially TOPAS, to perform accurate physical dose deposition simulations in carbon-ion therapy.
\begin{figure} [!htbp]
	\centering
	\includegraphics[width=0.75\columnwidth]{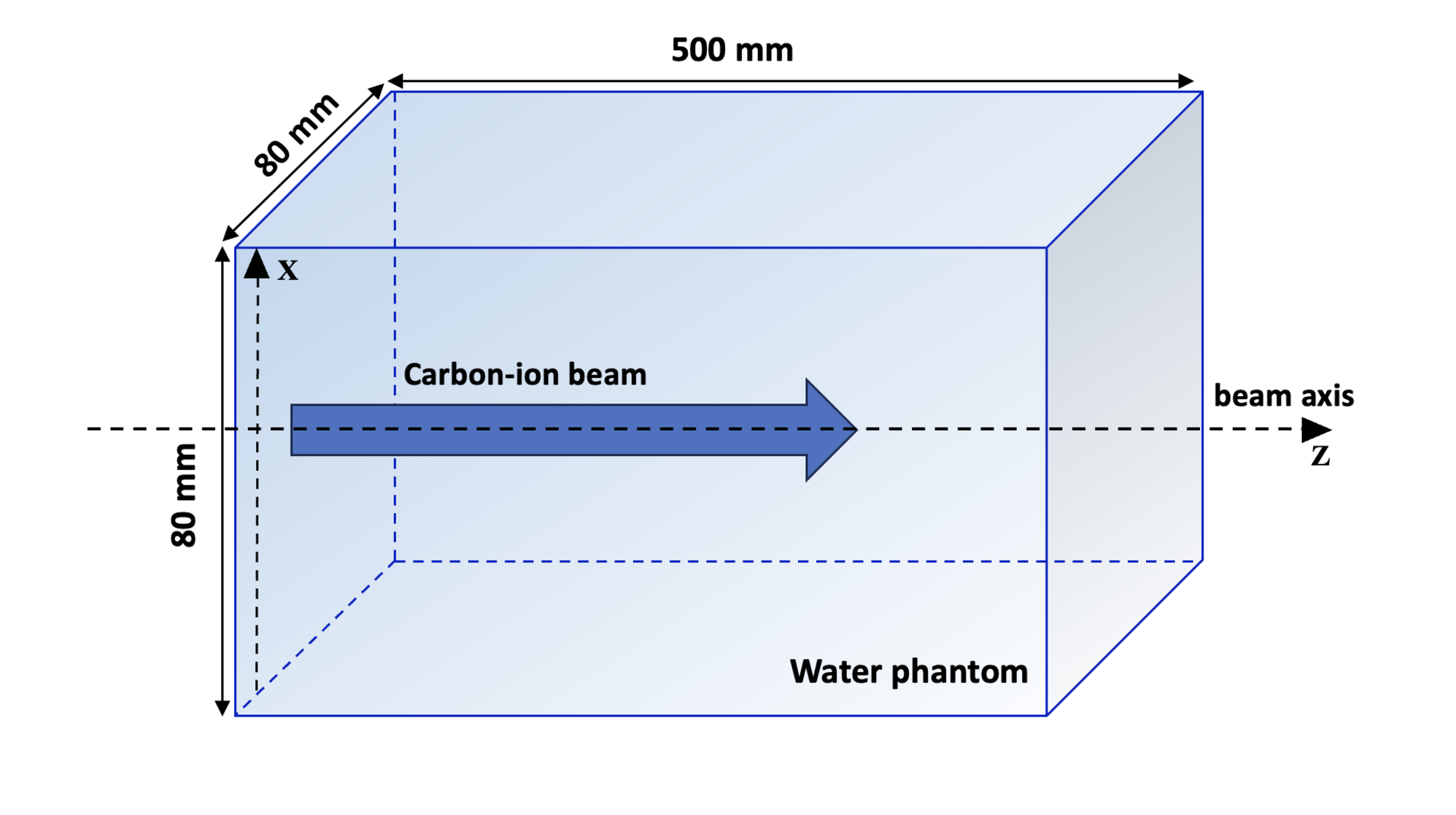}
	\caption{Topas Simulation Setup}
	\label{fig:topas_sim}
\end{figure}

In this study, the depth-dose distributions of mono-energetic carbon-ion beams were simulated for 280 and 430 MeV/u energies using TOPAS version 3.8.p.1 with Geant4 version 10.7. Figure 2 shows the geometry of the setup. The simulation configuration includes an 80 × 80 × 500 mm³ water phantom and the dose is scored along the beam axis with a voxel resolution of 1 × 1 × 1 mm³ (Halıcılar \textit{et al} 2024). For each energy, $10^{6}$ carbon ions were simulated corresponding to the statistical configurations used in similar MC-based studies to ensure maximum estimation of Bragg peak behavior and contributions from secondary fragments (Francis \textit{et al} 2014, Ying \textit{et al} 2019, El Bekkouri \textit{et al} 2023). The physics list used in TOPAS contains some of the necessary modules for high-precision simulation of ion beams: ‘g4em-standard-opt4’ and ‘g4em-extra’ for electromagnetic interactions, ‘g4ion-binary cascade’ for inelastic ion-nuclear collisions, ‘g4h-elastic-HP’ and ‘g4stopping’ for high-precision calculations of elastic processes involving hadrons, and ‘g4h-phy-QGSP-BIC-HP’ for inelastic nuclear interactions. Moreover, radioactive and particle decay processes are accounted for using ‘g4radio-activedecay’ and ‘g4decay’. This comprehensive configuration enables precise simulation of the Bragg peak and distal fragmentation, the two most critical parameters for treatment planning in carbon-ion radiotherapy.

\section{Results}

The depth-dose curves of carbon-ion beams at different energies were simulated using the proposed analytical model in a MATLAB computational environment. Figure 3 displays the depth-dose distribution profiles of carbon-ion beams for energy ranges of 100 MeV/u to 400 MeV/u. This confirms the fundamental physical characteristics of carbon-ion interactions with materials that are equivalent to water. These results concur with those of earlier research by Schardt \textit{et al} (2010).

As the beam energy rises from 100 MeV/u to 400 MeV/u, Figure 3 shows that the Bragg peak moves progressively deeper within the medium. The penetration depth specifically varies between 23 mm at 100 MeV/u and 276 mm at 400 MeV/u. This demonstrates the significant energy dependency of ion penetration in water-equivalent materials and more than a twelvefold increase in range (Kanai et al 1999, Schardt et al 2010). At the same time, the relative entrance dose, expressed as a percentage of the peak dose, decreases substantially from 65\% at 100 MeV/u to less than 20\% at 400 MeV/u, reflecting a significant increase of normal tissue sparing directly adjacent to the target volume (Kraft 2000, Durante and Paganetti 2016).
\begin{figure} [!htbp]
	\centering
	\includegraphics[width=0.75\columnwidth]{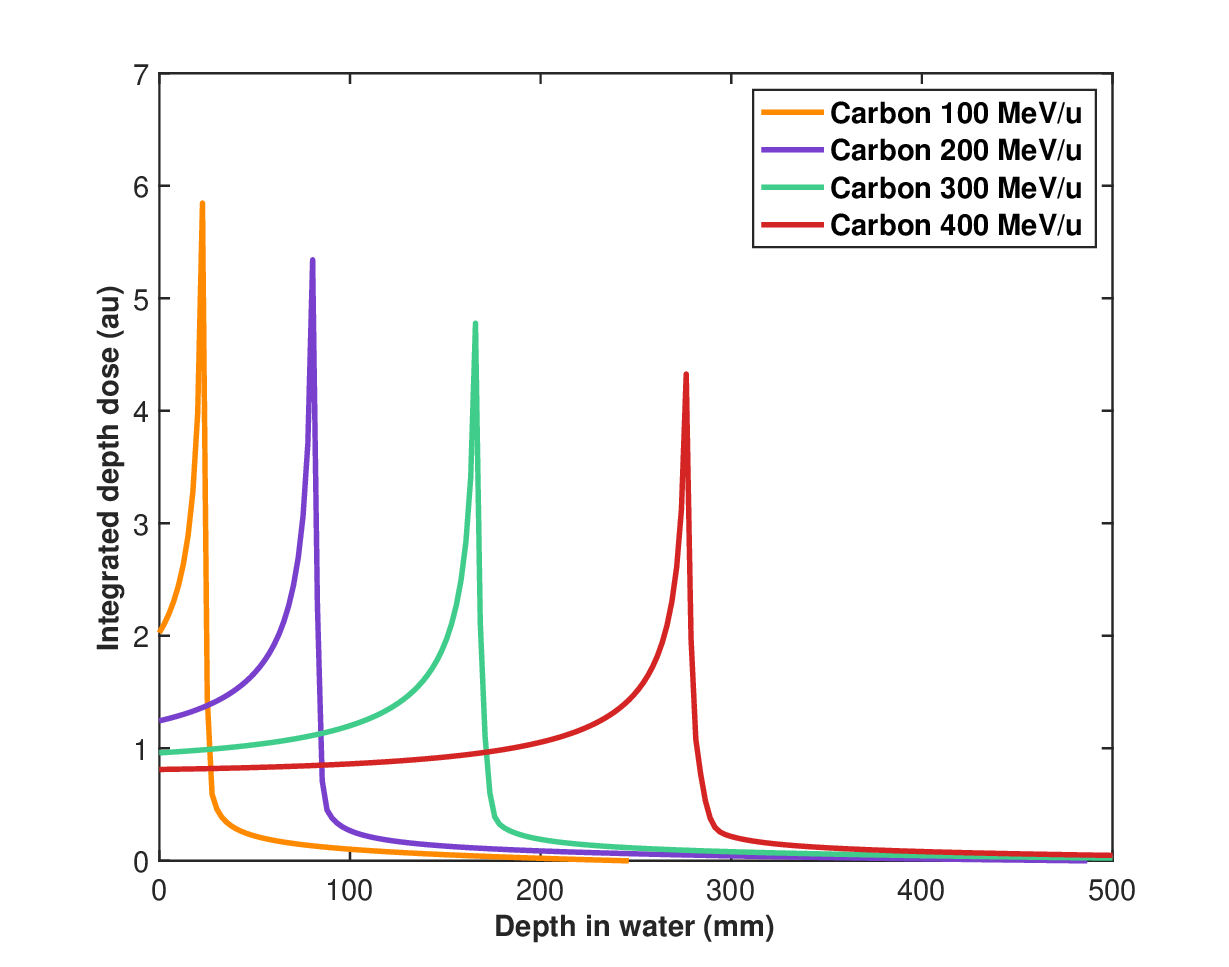}
	\caption{The dose distribution for carbon-ion beams with energies of 100 MeV/u, 200 MeV/u, 300 MeV/u, and 400 MeV/u.}
	\label{fig:carbon100_400 }
\end{figure}

The 100 MeV/u beam (orange curve) shows a distinct Bragg peak at 23 mm with a peak dose of 5.9 a.u., which is ideally used for superficial tumor therapy and pediatric tumors where shallow penetration and accurate localization are essential (Kamada \textit{et al} 2015). The 200 MeV/u beam (purple curve) reaches 80 mm with a maximum of 5.4 arbitrary units. This corresponds to a 3.6-fold range increase with a comparable peak amplitude dose and hence is appropriate for mid-range tumor positions, for instance, head and neck tumors (Schulz-Ertner and Tsujii, 2007). At higher energies, the 300 MeV/u beam (green curve) goes up to 166 mm with a peak dose of 4.9 a.u., whereas the 400 MeV/u beam (red curve) penetrates to 276 mm with a peak dose of 4.4 arbitrary units. The gradually decreasing peak dose amplitude with increasing energy illustrates decreasing linear energy transfer with increasing particle velocities (Scholz \textit{et al} 1997). These energies are best suited for the treatment of tumors located deeply in the abdominal, pelvic, and thoracic cancers (Durante and Loeffler, 2010).

Notably, the plateau of the entrance dose remains consistently low at all energies, at around 1 a.u., and demonstrates the superb sparing of normal tissues by carbon-ion beams compared to standard photon radiotherapy. Moreover, the Bragg peak broadens with increasing energy, which is given by the increase in full width at half maximum (FWHM) of approximately 85 \%, and this can be utilized in the volume tumor coverage, but needs to be planned to avoid exposure of the surrounding normal tissue. The distal dose gradient also reduces with high energy, resulting in more gradual fall-off to shape dose conformity to critical structures (Inaniwa  \textit{et al} 2015b).

\begin{figure}[!htbp]
	\centering
	\includegraphics[width=0.75\columnwidth]{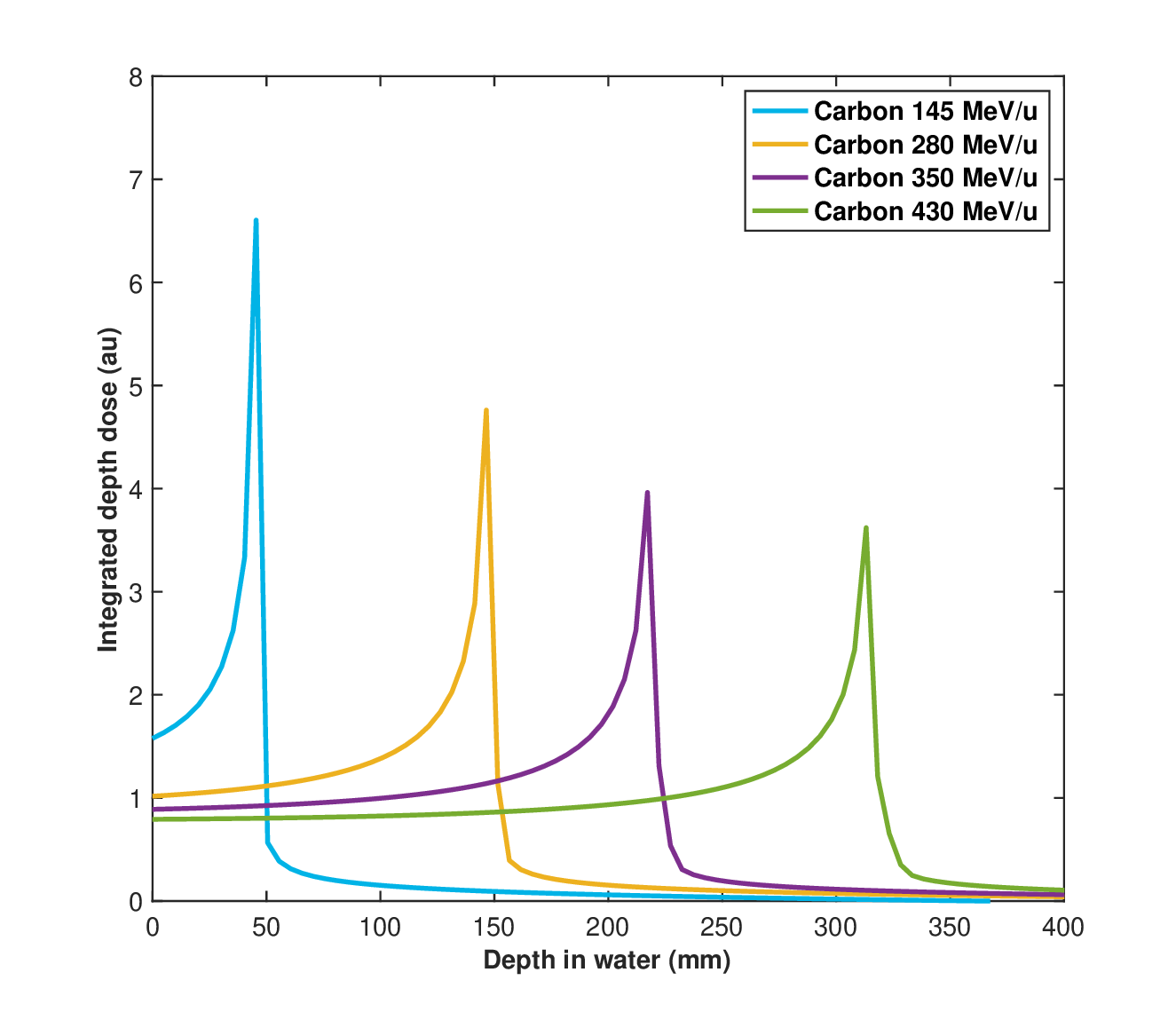}
	\caption{The depth-dose distribution of carbon-ions, calculated using the proposed analytical model for carbon energies between 145 and 430 MeV/u.}
	\label{fig: carbon dose distribution }
\end{figure}

Figure 4 gives depth–dose distributions in water for monoenergetic carbon-ion beams of energies 145, 280, 350, and 430 MeV/u computed using the proposed analytical model. Each curve shows Bragg peak effect behavior of heavy charged particles with relatively small entrance dose, steep increase in dose deposition towards particle range end, and fast drop-off after the peak. The height and position of the Bragg peak are altered with the increase in beam energy, accounting for energy dependence of dose deposition profiles and range in carbon-ion therapy (Scholz and Kraft 1996, Kanai \textit{et al} 1999, Kraft, 2000).

For the 145 MeV/u beam, the Bragg peak appears at a depth of 45 mm with a maximum dose of 6.6 arbitrary units. This curve has a steep dose gradient and a sharply localized peak, indicating minimal nuclear fragmentation and high linear energy transfer (LET). It is especially well-suited for the high precision and conformality treatment of superficial cancers because of these characteristics (Furusawa \textit{et al} 2000, Suzuki \textit{et al} 2000). In comparison to the 145 MeV/u beam, the 280 MeV/u beam had a 27 \% lower peak amplitude and reached its Bragg peak at 146 mm with a peak dose of 4.8 a.u. The entrance-to-peak amplitude difference reflects increased energy deposition over a longer path, accompanied by the onset of moderate nuclear fragmentation. There is a wider plateau before the peak and a modest distal tail becomes apparent (Kanai \textit{et al} 1999, Pshenichnov \textit{et al} 2008). For the 350 MeV/u beam, the Bragg peak is located at 217 mm with a reduced peak dose of 4.0 arbitrary units. The extended fragmentation tail after the Bragg peak is more noticeable, although the sharp decrease at the distal end persists. This aligns with higher-energy ion interactions and indicates a gradual decline in LET with increasing beam energy (Krämer and Scholz, 2000; Tsuji \textit{et al} 2005). At 430 MeV/u, the Bragg peak is found at 313 mm, and the maximum dose decreases further to 3.6 a.u. At this energy, nuclear fragmentation is more extensive, producing a broader tail and reducing the overall spatial confinement of the peak (Inaniwa \textit{et al} 2010).

Due to the cumulative contributions from multiple Coulomb scattering and nuclear interactions, the plateau region preceding the Bragg peak becomes increasingly noticeable at higher energies (Pshenichnov \textit{et al} 2008). The distal tail dose beyond the Bragg peak becomes increasingly evident due to increased fragment production at energies greater than 300 MeV/u (Inaniwa \textit{et al} 2010). The analytical model can precisely reproduce the asymmetric shape of the Bragg peak, which is defined by a gradual rise in dose up to the peak and a rapid decline afterward. The region of steep dose decrease beyond the peak, quantified as the distance over which the dose falls from 80\% to 20\% of its maximum, consistently ranges between 10 and 15 mm for all beam energies studied. This result aligns with clinical requirements for high-precision dose delivery in carbon-ion radiotherapy and confirms the model's ability to reproduce the essential physical features of carbon-ion interactions in water.

Figure 5 shows normalized integral depth–dose distributions (IDDDs) for 145, 280, 350, and 430 MeV/u monoenergetic carbon-ion beams in water. The dose profiles normalized to their respective Bragg peak maxima demonstrate the expected energy-dependent shift of the peak locations at approximately 46, 147, 217, and 313 mm, respectively. This is consistent with established energy–range relations of carbon ions in water, which supports the spatial precision of the analytical model.
\newpage
The entrance dose rises with increasing beam energy, from about 21\% of the dose at 145 MeV/u to about 33\% at 430 MeV/u. This is due to the enhanced secondary particle contribution and reduced linear energy transfer (LET) with increasing energies. The distal fall-off becomes less steep with increasing energy, showing increased nuclear fragmentation and the production of lighter secondary fragments delivering dose beyond the Bragg peak. 

These analytical calculations are in agreement with MonteRay results (Lysakovski \textit{et al} 2024). MonteRay simulations provided Bragg peak depths of around 46, 148, and 315 mm for 150, 280, and 430 MeV/u beams, respectively, to be in reasonable agreement with analytical model predictions. The entrance-to-peak dose ratios and distal tail behavior seen in MonteRay agree with the results obtained using the analytical method.

\begin{figure} [htbp]
	\centering
	\includegraphics[width=0.75\columnwidth]{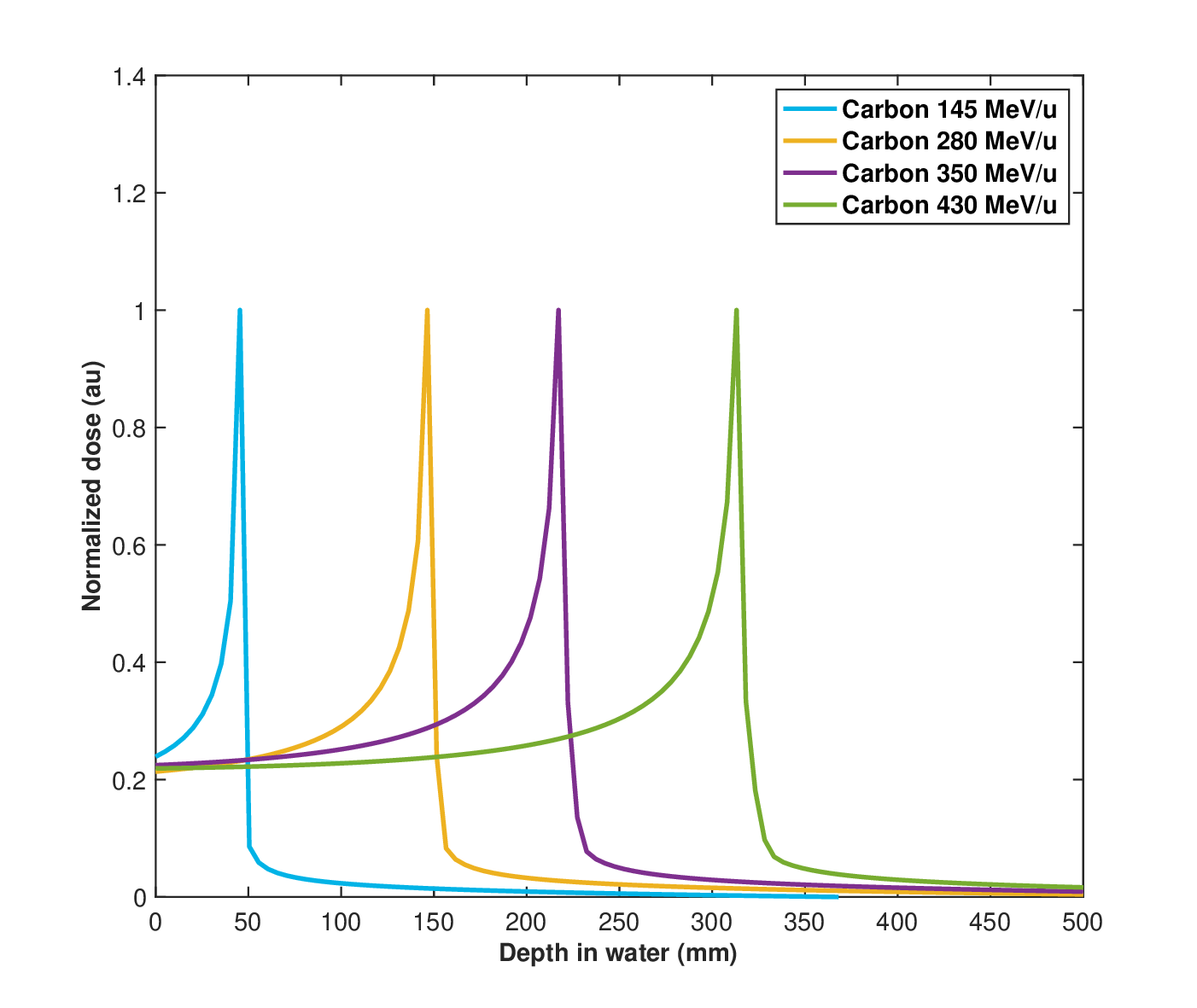}
	\caption{Normalized depth dose distributions (IDDDs) with a range of 145 to 430 MeV/u for quasi-monoenergetic carbon-ion beams.}
	\label{fig:Normalized_dose }
\end{figure}
The agreement between the proposed analytical model and MonteRay simulations reflects the reliability of the proposed model for rapid dose prediction in carbon ion radiotherapy. The analytical model provides a complementary tool for preliminary treatment planning and verification, enabling accurate and efficient dose assessments in clinical processes, while MonteRay enables high-fidelity dose calculations with significant computational efficiency.

Figure 6 shows depth–dose curves in water for carbon-ion beams with energies 280 MeV/u and 430 MeV/u, which are calculated by the proposed analytical model. Bragg peaks are indicated around 147 mm in the case of the 280 MeV/u beam and 313 mm in the case of the 430 MeV/u beam, in agreement with theoretical energy–range relations for carbon ions in water. With increasing beam energy, the entrance dose becomes more prominent as a result of decreased linear energy transfer (LET) and greater secondary particle contribution. The 430 MeV/u beam also displays a less steep and more gradual distal fall-off than the 280 MeV/u beam, corresponding to greater nuclear fragmentation and increased range of lighter secondary fragments beyond the Bragg peak. These findings confirm that the analytical model successfully simulates the major physical properties of high-energy carbon-ion interactions and is found to be competent for successful and clinically relevant dose calculations in heavy-ion radiotherapy.

\begin{figure} [htpt!]
	\centering
	\includegraphics[width=0.75\columnwidth]{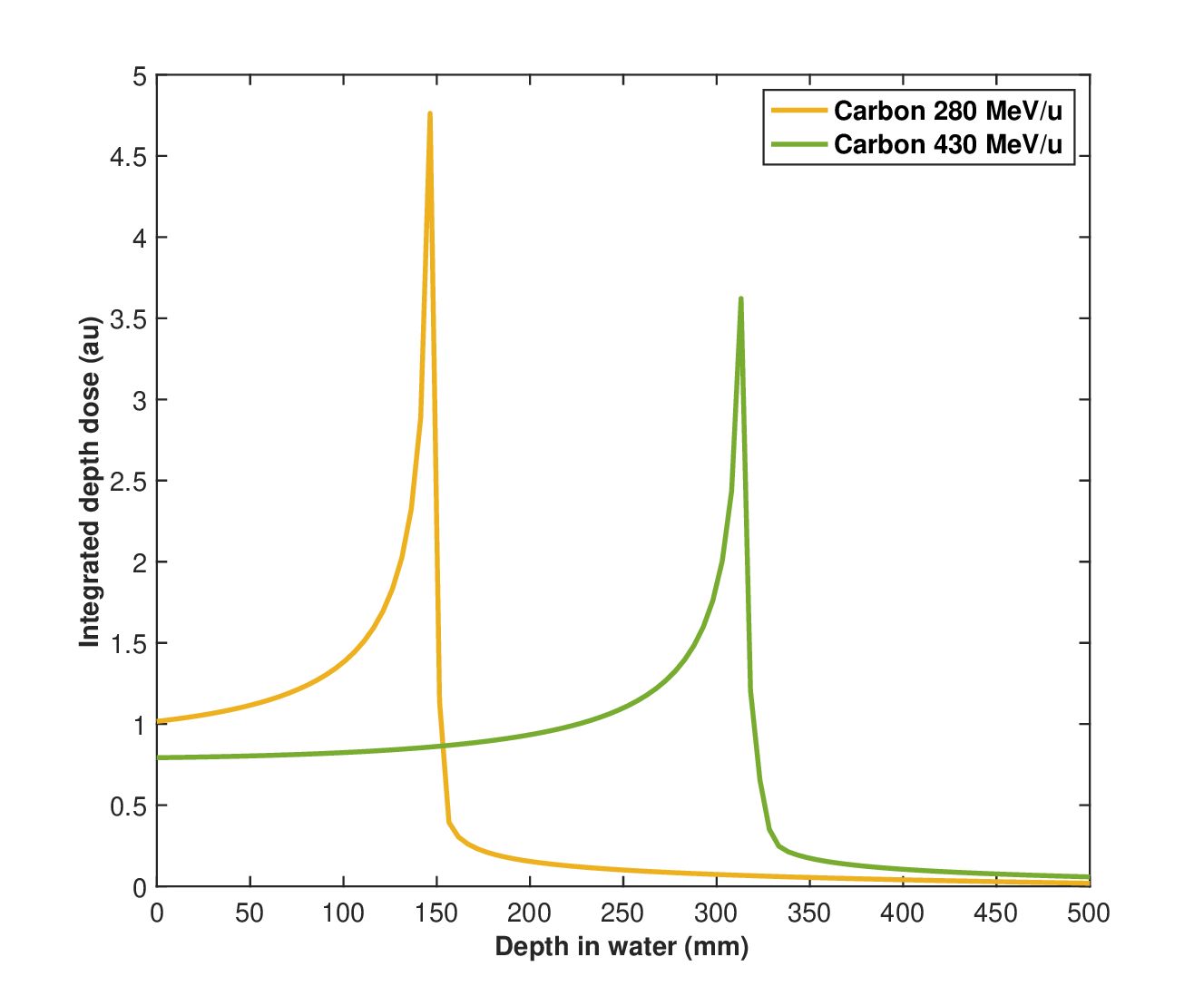}
	\caption{Dose distribution of carbon-ion for beams with 280 MeV/u and 430 MeV/u in water. }
	\label{fig:Car280_430 }
\end{figure}

\subsection*{Comparison of Analytical Model with TOPAS Simulations}

Figure 7 shows the comparison between the calculated depth-dose distribution from the analytical model and the Monte Carlo simulation outcome of TOPAS for a carbon-ion beam with an energy of 280 MeV/u. The Bragg peak positions agree well with the peak at 147 millimeters by the analytical model and 153 millimeters by the TOPAS simulation. This induces an absolute range underestimation of 6.5 millimeters. This discrepancy indicates that the stopping power equation used in the analytical model does not accurately cover intricate energy-loss processes that become more significant as the ions decelerate close to the Bragg peak. Specifically, electron shell effects and medium-dependent density corrections, both incorporated in Monte Carlo simulations, may be underestimated in the analytical model. 

\begin{figure} [htpt]
	\centering
	\includegraphics[width=0.75\columnwidth]{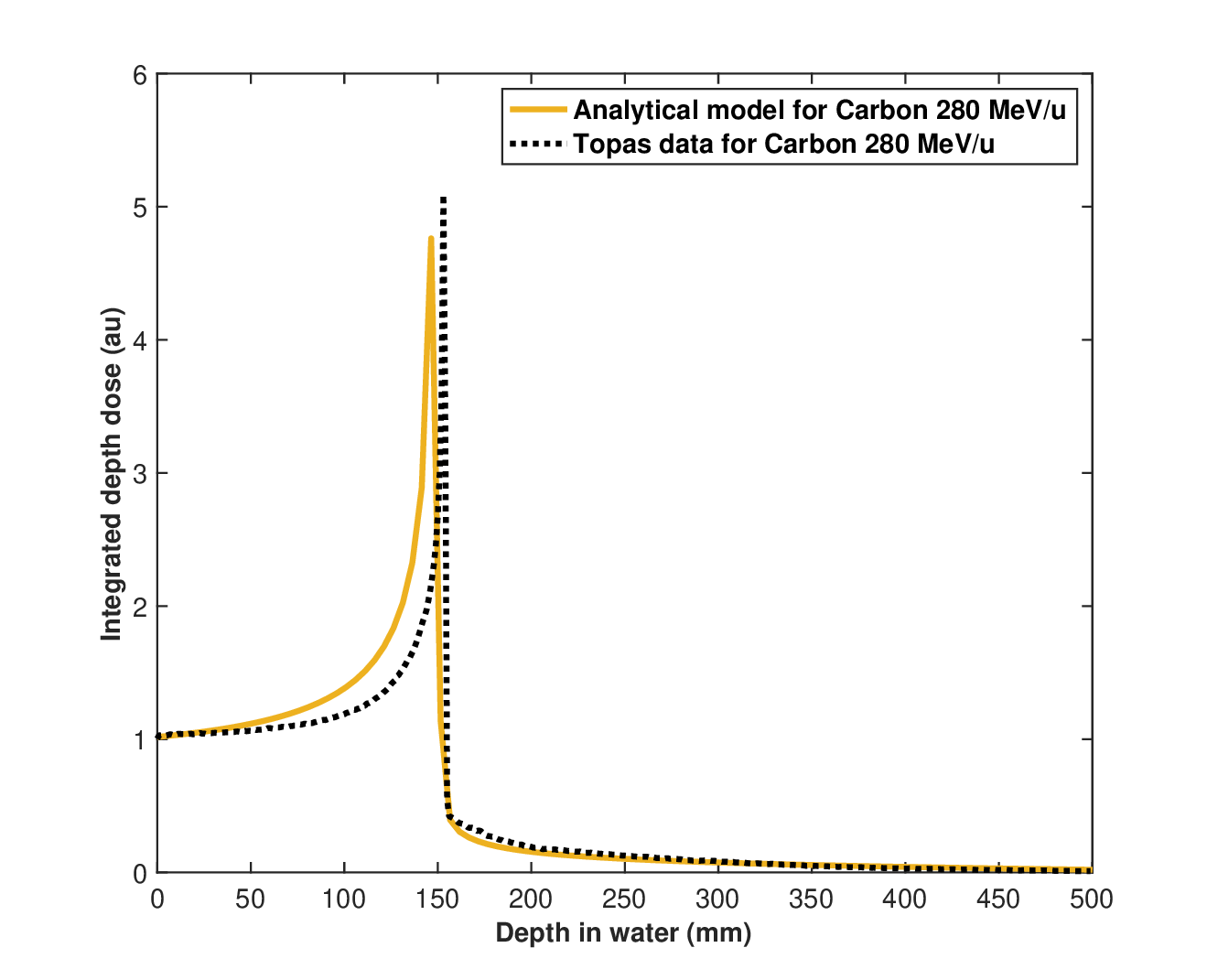}
	\caption{Comparison of analytical models and TOPAS simulation for the integrated depth dose curve of a 280 MeV/u carbon-ion beam.}
	\label{fig:Car280 }
\end{figure}

The analytical model also indicates good agreement in the plateau region, with dose variations within 5–7\% of the Monte Carlo results. This agreement demonstrates the model's success in simulating pre-peak energy deposition, where electronic interactions predominate and can be handled analytically. However, the peak dose amplitude in the analytical calculation is around 6\% lower than that predicted by TOPAS simulation. This difference most probably stems from the simplification of the nuclear fragmentation and energy straggling processes in the analytical approach, which become the dominant process at energies close to 280 MeV/u. Compared to TOPAS simulation, which includes detailed modeling of projectile and target fragmentation as well as electromagnetic dissociation, the analytical model relies on generalized approximations that may reduce its accuracy in the high-dose region near the Bragg peak. Besides, the post-peak region shows a significantly steeper dose falloff in the analytical analysis than the broad distal tail modeled by TOPAS. This is due to the poor capacity of the analytical model to simulate the transport and energy deposition of the recoil nuclei and secondary fragments that contribute to residual dose beyond the Bragg peak. Despite these systematic differences, the total dose difference is less than 8\% across the entire profile, showing that the analytical model can provide clinically acceptable accuracy for most treatment planning uses while providing considerable computational benefit compared to complete Monte Carlo simulations.

Figure 8 presents the comparison of depth–dose distributions between the analytical model and the TOPAS Monte Carlo simulation for a carbon-ion beam with an energy of 430 MeV/u. The positions of the Bragg peaks are in relatively close agreement, at a peak of 313 millimeters from the analytical model and 308 millimeters from the TOPAS simulation. This is a range overestimation of 5 millimeters. The agreement in range indicates that the analytical model's stopping power formulation performs reasonably well at higher energies. However, some divergence remains based on simplifying assumptions regarding ion deceleration and energy loss processes in water.

\begin{figure} [htpt]
	\centering
	\includegraphics[width=0.75\columnwidth]{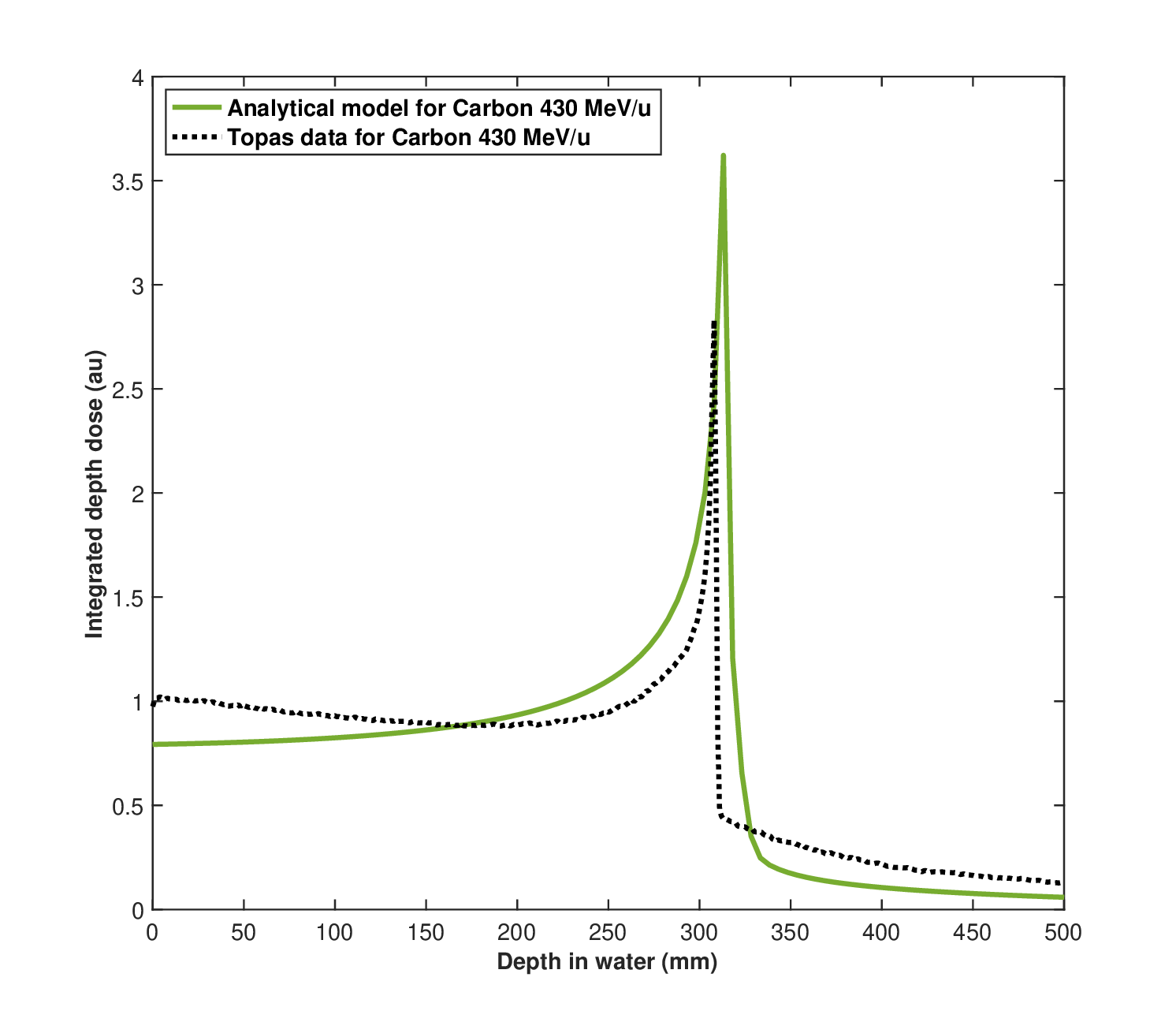}
	\caption{Analytical model and TOPAS simulation comparison for the integrated depth dose profile of a 430 MeV/u carbon-ion beam}
	\label{fig:Car430 }
\end{figure}

Although there is positional agreement, significant differences in dose magnitude are observed, especially in the plateau and entrance regions. The analytical model overestimates the entrance dose by approximately 10\% compared to TOPAS, and the plateau region shows deviations up to 12\%. These overestimations may result from limitations in the analytical model's handling of lateral scattering and relativistic energy loss mechanisms, which become more complex at higher energies. The amplitude of the Bragg peak estimated by the analytical model is also approximately 15 \% lower than that of the Monte Carlo simulation. This underestimation likely stems from the model’s simplified handling of inelastic nuclear interactions like projectile fragmentation and cumulative energy straggling that are modeled more extensively within Monte Carlo frameworks.

\newpage

A more pronounced discrepancy occurs in the distal fall-off region after the Bragg peak. The analytical model demonstrates a steeper dose decline, whereas the TOPAS simulation shows a longer distal tail, resulting from the ongoing energy deposition by recoil nuclei and secondary fragments. The analytical model underestimates this component by approximately 10–15 \%, reflecting its limited ability to simulate the complex transport and interaction of heavy secondary particles. These findings underscore the need to enhance the analytical model's capacity to include secondary particle transport and fragmentation physics, especially at higher energy levels. However, the overall profile remains within a reasonable limit of clinical acceptability, which implies that the model can be suitably applied in high-energy scenarios where rapid dose estimation and computational efficiency are prioritized.

Table 1 shows a comparison of the carbon-ion range estimates obtained from the analytical model and TOPAS Monte Carlo simulation at beam energies of 280 MeV/u and 430 MeV/u. For 280 MeV/u, the analytical model's range prediction is 146 millimeters, while the TOPAS simulation gives a range of 153 millimeters, which is an underestimation of approximately 4\%. This discrepancy is likely caused by the simplified treatment of energy loss mechanisms and nuclear interactions in the analytical model, which become increasingly significant at intermediate energies. On the other hand, the analytical model overestimates the range at 430 MeV/u, estimating 313 millimeters compared to 308 millimeters from the TOPAS simulation, which is a discrepancy of approximately 2\%. This closer alignment at higher energy indicates enhanced model performance in conditions where nuclear fragmentation is comparatively less significant and electronic stopping power predominates. According to these results, the analytical model maintains a clinically relevant energy spectrum with acceptable accuracy, despite small systematic variations. Additionally, it supports its potential use in scenarios requiring rapid and computationally efficient dose estimation in carbon-ion radiotherapy.

\begin{table}[htpt]
\caption{Comparison of carbon-ion range estimates using analytical model and TOPAS simulation at beam energies 280 MeV/u and 430 MeV/u.}
\vspace{0.5 cm} 
\centering
\begin{tabular}{|c|c|c|}
\hline
 {\textbf{\shortstack{Energy\\(MeV/u)}}} & {\textbf{\shortstack{Analytical Model\\Range (mm)}}} & \multicolumn{1}{c|}{\textbf{\shortstack{TOPAS\\Range (mm)}}} \\
\hline
280 & 146 & 153 \\
\hline
430 & 313 & 308 \\
\hline
\end{tabular}
\label{tab:crb_ranges}
\end{table}

\section{Discussion}
Precise dose calculation remains a fundamental requirement in carbon ion radiotherapy (CIRT), a treatment modality with clinically relevant advantages because of its superior dose conformity and higher relative biological effectiveness compared to conventional photon radiotherapy (Loeffler and Durante, 2013). These features are particularly valuable for treating radioresistant tumors situated in deep-seated regions. However, the complex physical interactions of carbon ions, such as nuclear fragmentation, lateral scattering, and highly localized deposition, present substantial obstacles to precise dose prediction and lead to computationally challenging calculations. Although Monte Carlo simulations provide highly precise dose calculations, their long computation times restrict routine clinical application (Parodi \textit{et al} 2012). Consequently, analytical dose calculation models have become fundamental elements in CIRT, allowing fast and reliable dose predictions that can be integrated into real-time treatment planning systems (Inaniwa \textit{et al} 2014). The improvement and development of these models have a direct impact on the reliability of dose prediction, the estimation of biological effects, and the clinical safety of CIRT.

However, accurate dose calculation in CIRT presents several related challenges. The first major challenge concerns modeling the physical dose distribution, which is influenced by lateral scattering, the Bragg peak, and the fragmentation tail beyond the target volume. Analytical pencil beam models are prevalently used to simplify the complex dose deposition by approximating the lateral dose spread around every pencil beam path with Gaussian distributions. To enhance clinical accuracy, Parodi \textit{et al} (2013) developed a Monte Carlo-based lateral dose spread parameterization for scanned proton and carbon ion beams, which improved on the pencil beam method. Zhang \textit{et al} (2018) derived a novel pencil beam model directly from Monte Carlo simulations to improve dose calculation accuracy while maintaining clinical efficiency. More recently, Yagi \textit{et al} (2022) implemented a fast-raster scanning pencil beam algorithm optimized for clinical carbon ion delivery in the VQA Plan system. Nevertheless, conventional pencil beam models are based on a single Gaussian approximation and may fall short in heterogeneous tissues or distal regions affected by nuclear fragmentation. To overcome these limitations, Inaniwa \textit{et al} (2014) introduced the triple Gaussian beam model to enhance the separation of the lateral dose into three terms of Gaussians representing primary carbon ion and nuclear fragment contributions. This model greatly enhances dose prediction within off-axis and heterogeneous areas. This concept was extended biologically by Inaniwa and Kanematsu (2015) in their trichrome beam model, which improved the precision of biological dose predictions in clinical treatment planning.

A second significant obstacle is the precise estimation of the biological dose, which depends on the relative biological effectiveness (RBE), which changes with depth, linear energy transfer (LET), and tissue type. Several radiobiological models have been developed to handle this complexity. The Local Effect Model (LEM), first proposed by Scholz and Kraft (1996) and developed in subsequent research (Scholz \textit{et al} 1997, Kraft \textit{et al} 1999, Kramer \textit{et al} 2000, Grün \textit{et al} 2012), predicts relative biological effectiveness (RBE) by linking photon reference data to local energy deposition and track structure of heavy charged particles. Conversely, the Microdosimetric Kinetic Model (MKM) was first developed by Kase \textit{et al} (2006) and later used in clinical practice by Inaniwa \textit{et al} (2010). It applies microdosimetric parameters and models of stochastic energy deposition to estimate cell-killing probabilities. It has been clinically evaluated recently by Fossati \textit{et al} (2018).  Complementing these models, the mixed beam model first proposed by Kanai \textit{et al}  (1997) separates the biological effects of primary carbon ions, heavy nuclear fragments, and light secondary particles. Inaniwa \textit{et al}  (2015b) reformulated and clinically applied this model to increase robustness and accuracy for treatment planning. Karger and Peschke (2018) present a thorough review of progress in carbon ion RBE models for carbon ion therapy. These models form a strong basis for biologically optimized treatment planning in CIRT, enabling better therapeutic efficiency and patient safety.

Several experimental studies have been carried out on the development and validation of carbon-ion dose distribution models. Foundational experimental research by Haettner \textit{et al} (2013) rigorously established the physical properties of carbon-ion beams, such as energy deposition and nuclear fragmentation in water, and delivered essential empirical measurements to subsequent computational models. Zhang \textit{et al} (2022b) also validated a novel pencil-beam model derived from Monte Carlo calculations to exhibit good agreement between theoretical predictions and measured dose distributions for a wide range of clinical cases. Yagi \textit{et al} (2023) conducted thorough experimental verification of radiobiological optimization approaches under conditions of intensity-modulated carbon-ion therapy, while Fujitaka \textit{et al} (2021) provided direct experimental explanation of physical and biological beam characteristics in scanning beam delivery systems. Yogo \textit{et al} (2021) pioneered the high-speed and high-resolution ZnS scintillator sheet-based dose distribution measurement in three dimensions, followed by LET-based analysis by Nakaji \textit{et al} (2022). Together, these experiments constitute an integrated validation framework that enhances the physical accuracy and clinical practicability of modern carbon-ion dose calculation models.

Clinical carbon-ion beam energies have been thoroughly studied in the literature for tumor treatment at various depths, with energies chosen to place the Bragg peak at the tumor site and maximize biological effectiveness. Low energies like 95 MeV/u have been investigated for superficial tumors using experimental measurements of double-differential fragmentation cross-sections on thin targets to present data for modeling secondary particle production (Dudouet \textit{et al} 2013). At 140 MeV/u, 240 MeV/u, and 344 MeV/u, a novel pencil beam model for carbon-ion dose calculation derived from Monte Carlo simulations was constructed to enhance the precision of dose calculation at these clinical energies (Zhang \textit{et al} 2018). For 195 MeV/u, semi-analytical radiobiological models were also introduced to support biologically informed treatment planning (Kundrát 2007).  Physical beam modeling with dose optimization frameworks was also developed for this energy (Krämer \textit{et al} 2000). At 270 MeV/u, quite a number of studies examined various models such as: radiobiological modeling (Kundrát 2007), optimized beam models (Krämer \textit{et al} 2000), benefits of heavy-ion beams regarding depth-dose characteristics and biological effectiveness (Schardt \textit{et al} 2010), and simulated Bragg curves using GEANT4 (Hamad 2021). For 290 MeV/u, the dosimetric precision of nuclear interaction models in Geant4 was examined (Kameoka \textit{et al} 2008), and Monte Carlo simulations characterized mixed radiation fields and secondary fragmentation effects in water phantoms (Ying \textit{et al} 2017, 2019). For 330 MeV/u beams, experimental data on nuclear fragmentation in water (Haettner \textit{et al} 2013) and semi-analytical models (Kundrát 2007) were utilized to improve predictive models and verify the accuracy of simulations. These discrete energy levels represent a systematic clinical plan for adjusting carbon-ion beams to various anatomical regions and tumor depths.

The energy levels of clinical carbon-ion beams usually vary from 100 to 430 MeV/u, depending on tumor depth and treatment requirements (Yoon \textit{et al} 2023). The proposed analytical model for dose calculation, implemented in MATLAB, focuses on this energy range to enhance the precision of treatment planning. A critical feature of carbon-ion therapy dose distributions is the fragmentation tail appearing both before and after the Bragg peak, caused by nuclear interactions between primary ions and tissue nuclei that generate secondary fragments such as protons, helium, lithium, beryllium, boron, and heavier isotopes like $^{11}\mathrm{B}$ and  $^{11}\mathrm{C}$ in front of the Bragg peak and  $^{10}\mathrm{B}$ beyond the peak (Nandy 2021). Primary ion scattering and energy loss cause the tail before the Bragg peak, while secondary fragments moving deeper into healthy tissue are mostly responsible for the tail after the peak (Günzert-Marx \textit{et al} 2008, Ying \textit{et al} 2017). Protons and helium ions are the dominant contributors beyond the peak due to their abundance and longer ranges (Matsufuji \textit{et al} 2003, Günzert-Marx \textit{et al} 2008), while lithium, beryllium, and boron fragments also impact dose and biological effects (Schardt \textit{et al} 2010). Heavy fragments affect dose build-up and require extensive modeling for precise dosimetry and optimizing relative biological effectiveness (Parodi \textit{et al} 2012). In this research, we focus on  $^{11}\mathrm{B}$ and  $^{11}\mathrm{C}$ fragments that occur before the Bragg peak,  $^{10}\mathrm{B}$ fragments that appear after the Bragg peak, and light secondary particles such as protons and helium ions appearing beyond the Bragg peak to reflect their respective contributions to the fragmentation tail in the analytical model.

The depth–dose distributions of the 100, 200, 300, and 400 MeV/u carbon-ion beams in water are illustrated in Figure 3, which were calculated by the proposed analytical model in MATLAB. The Bragg peak positions were measured to be 23 mm for 100 MeV/u, 80 mm for 200 MeV/u, 166 mm for 300 MeV/u, and 276 mm for 400 MeV/u. These findings align well with earlier published experimental results and Monte Carlo simulation data, reinforcing the reliability and robustness of our model. At 200 MeV/u, our estimated Bragg peak of 80 mm is in good agreement with several recognized references. Haettner \textit{et al} (2013) experimentally identified the Bragg peak at approximately 80-82 mm, demonstrating the effects of nuclear fragmentation that contribute to the dose tail after the Bragg peak. Schardt et al. (2010) also observed the Bragg peak at around 80 mm in their extensive review of heavy-ion physics. This finding was validated by Liu (2017) through advanced Monte Carlo simulations with TOPAS, supporting the precision of our analytical model within a 1–2 mm margin. Our analytical model predicts a Bragg peak at 166 mm for a 300 MeV/u carbon-ion beam, which aligns closely with the findings of Schardt \textit{et al} (2010), who reported a range of approximately 165-167 mm. Liu (2017) also confirmed this range using comprehensive TOPAS Monte Carlo simulations, demonstrating that carbon ions at 300 MeV/u show not only the predicted penetration depth but also a distinct narrow lateral spread, which is crucial for attaining high-dose conformity. At this energy, our model precisely represents the extended fragmentation tail that occurs after the Bragg peak, which is essential for accurate dose calculations in treatment planning systems. For 400 MeV/u, our model computed range of 276.4 mm shows remarkable agreement with significant experimental and simulation research. El Bekkouri \textit{et al} (2023) indicated a peak of 274.2 mm through PHITS simulations that considered the contributions of secondary particles, whereas Haettner \textit{et al} (2013) obtained a peak of 274.6 mm experimentally. Additionally, Kramer \textit{et al} (2000) determined a Bragg peak at 276 mm through a physical beam model for treatment planning, which validates our findings with a discrepancy of less than 1\%. These results validate that our analytical model accurately simulates carbon ion depth-dose distributions throughout the entire clinically significant energy range, reflecting both the beneficial characteristics of the Bragg peak and the complex contributions from secondary fragments. In carbon ion radiotherapy treatment planning, this method can provide a computationally efficient option for rapid Monte Carlo simulations.

The Bragg curves for carbon-ion beams at energies of 145, 280, 350, and 430 MeV/u in water are depicted in Figure 4, computed by the proposed analytical model in MATLAB. The positions of the Bragg peaks were determined to be 46, 147, 217, and 313 mm, respectively. These results illustrate the expected energy-dependent shift of the Bragg peak to greater depths and underscore the necessity of incorporating nuclear interaction effects in integrated depth–dose distributions for carbon-ion radiotherapy. At 280 MeV/u, the calculated Bragg peak depth of 147 mm is in close agreement with the experimental findings of Yang \textit{et al} (2021), who observed a depth between 145 and 147 mm using CR-39 detectors at an incident energy of 276.5 MeV/u. This result also agrees with the analytical findings of Kempe and Brahme (2010), who modeled carbon-ion beams at 279 MeV/u and reported similar peak positions. Likewise, the Bragg peak calculated at 217 mm for 350 MeV/u is consistent with the Monte Carlo simulations of Liu (2017) and the treatment planning results of Inaniwa \textit{et al} (2015a), both of which reported depths in the range of 215 to 218 mm. For 430 MeV/u, our model estimated the Bragg peak at 313 mm, which aligns well with the findings of Nakaji \textit{et al} (2022), who obtained a peak depth of approximately 312 to 314 mm, depending on LET-based measurements and Monte Carlo simulations used for clinical dose evolution. The overall consistency across all energies supports the accuracy of our proposed analytical approach in capturing complex physical processes like nuclear fragmentation and the production of secondary particles, which are essential for precise modeling of integrated dose distributions. Consequently, the analytical model provides a computationally efficient and adequately precise tool for both clinical treatment planning and research in carbon-ion radiotherapy.
\newpage
To assess the validity of the proposed analytical model, its calculated depth-dose distributions were compared with those obtained with TOPAS Monte Carlo simulations. Figure 7 illustrates a comparative analysis for a 280 MeV/u carbon-ion beam and shows strong agreement between our analytical model and the TOPAS simulations, especially in the Bragg peak location and the entrance region. However, several discrepancies were identified as the analytical model estimates a slightly higher dose in the plateau region before the Bragg peak, while the peak height is marginally lower compared to the TOPAS simulations. Furthermore, the falloff beyond the peak in the proposed analytical model is steeper, suggesting potential differences in modeling energy loss processes and secondary interactions. Figure 8 provides a similar comparison for the case of a 430 MeV/u carbon-ion beam, where the deviations between the proposed analytical model and the TOPAS simulations become more noticeable. Our analytical model estimates a lower entrance dose than that obtained from the TOPAS simulations, and the Bragg peak height is underestimated. Moreover, the fall-off after the Bragg peak shows noticeable disparity, suggesting that our model needs to be improved for higher energies.

The proposed analytical model demonstrates good performance in estimating depth-dose distributions at intermediate carbon-ion energies. However, it encounters notable limitations at higher energies due to its simplified treatment of complex physical interactions. One of the main limitations is the insufficient modeling of nuclear fragmentation and the generation of secondary particles, which undermines the ability of our model to accurately reproduce the distal dose tail observed in high-energy beam cases. This limitation is evident at 430 MeV/u, where the proposed analytical model overestimates the entrance and plateau region doses by about 10-12\% and underestimates the Bragg peak magnitude by about 15\% compared to TOPAS Monte Carlo simulations. These discrepancies indicate that our model inadequately represents inelastic nuclear interactions, energy straggling, and residual fragment transport, which become more pronounced at higher ion energies. Furthermore, the prediction of our model regarding a steeper dose falloff beyond the Bragg peak demonstrates a limited capacity to account for the extended dose contributions resulting from recoil nuclei and secondary fragments. While our model performs reasonably well at 280 MeV/u, its increasing divergence from results of Monte Carlo simulations at higher energies emphasizes the need to include more comprehensive physical modeling, such as lateral scattering, energy-dependent stopping powers and nuclear recoil effects for enhancing the stability and clinical utility of the analytical framework in high-energy carbon-ion therapy.

The proposed analytical model holds promising clinical potential by enabling rapid and reasonably accurate estimation of depth–dose distributions for carbon-ion beams across a broad energy spectrum. Its computational efficiency can give considerable advantages for preliminary treatment planning, energy selection, and adaptive radiotherapy workflows where time constraints are critical. At 280 MeV/u, which is a clinically significant energy for mid-depth tumor treatment, our model is in good agreement with the Monte Carlo simulation and accurately predicts the position of the Bragg peak and the overall dose profile. This performance demonstrates its suitability for planning situations requiring prompt and accurate dose evaluations. However, at higher energies, such as 430 MeV/u, the model's limitations are more apparent, especially due to the lack of proper modeling of nuclear fragmentation processes and secondary particle transport, which are important aspects for accurate dose calculation beyond the Bragg peak. This disparity may cause an underestimation of the distal dose tail, and this may compromise the safety of treatment in cases incorporating critical structures situated near or beyond the target volume. Furthermore, since the study assessed energies exceeding the typical clinical range of 100–300 MeV/u, further confirmation within this standard spectrum is pertinent. Consequently, while our analytical model can provide a strong balance of speed and performance, particularly in resource-limited environments, its clinical use should be approached judiciously and preferably supported by high-fidelity Monte Carlo simulations in complicated or high-risk treatment situations.

\section{Conclusion}

In conclusion, our analytical model can provide a computationally effective and reasonably accurate approach for carbon-ion beam depth-dose distribution estimation, indicating good consistency with TOPAS Monte Carlo simulations, especially in Bragg peak localization and general dose trends over clinically useful energies. Minor discrepancies seen in entrance dose levels, peak magnitude, and distal tail behavior are caused by the simplified treatment of nuclear interactions and the inadequate depiction of complicated fragmentation processes. In this work, our model accounts for significant secondary fragments, which comprise light ions like helium, protons and heavier nuclear fragments like $^{10}\mathrm{B}$ occurring beyond the Bragg peak and $^{11}\mathrm{B}$ and $^{11}\mathrm{C}$ contributing in front of the Bragg peak. The proposed analytical model enhances the dose deposition representation in the distal and proximal areas. Future research will concentrate on including a broader range of secondary fragments and improving the physical modeling of inelastic nuclear interactions to improve dosimetric accuracy, particularly in distal and lateral scattering zones. Furthermore, we plan to apply our model to heterogeneous media to allow personalized treatment planning in anatomically complicated patient geometries and to bridge the gap between analytical tractability and high-fidelity estimation provided by Monte Carlo-based systems.

\section{Acknowledgement}
We would like to thank Mehmet Burcin Unlu for his valuable suggestions concerning carbon-ion therapy and TUBA, the Turkish Academy of Sciences for financial support.

  \section*{References}

\noindent
\hangindent=2em
Abramowitz M and Stegun I A 1972 \textit{Handbook of Mathematical Functions} (New York: Dover)

\vspace{1em}
  
\noindent
\hangindent=2em
Agostinelli S \textit{et al} 2003 GEANT4—a simulation toolkit \textit{Nucl. Instrum. Methods Phys. Res. A} \textbf{506} 250--303

\vspace{1em}

\noindent
\hangindent=2em
Ahlen S P 1980 Theoretical and experimental aspects of the energy loss of relativistic heavily ionizing particles \textit{Rev. Mod. Phys.} \textbf{52} 121

\vspace{1em}

\noindent
\hangindent=2em
Allison J \textit{et al} 2016 Recent developments in Geant4 \textit{Nucl. Instrum. Methods Phys. Res. A} \textbf{835} 186--225

\vspace{1em}

\noindent
\hangindent=2em
Amaldi U and Kraft G 2005 Radiotherapy with beams of carbon ions \textit{Rep. Prog. Phys.} \textbf{68} 1861

\vspace{0.10em}

\noindent
\hangindent=2em
Aminafshar B, Baghani H R and Mowlavi A A 2024a Analytical parameterization of Bragg curves for proton beams in muscle, bone, and polymethylmethacrylate \textit{Radiol. Phys. Technol.} \textbf{17} 745--755

\vspace{1em}

\noindent
\hangindent=2em
Aminafshar B, Baghani H R and Mowlavi A A 2024b Analytical calculation of proton beam Bragg curve inside heterogeneous media \textit{Eur. Phys. J. Plus} \textbf{139} 1--12

\vspace{1em}

\noindent
\hangindent=2em
Battistoni G \textit{et al} 2015 Overview of the FLUKA code \textit{Ann. Nucl. Energy} \textbf{82} 10--18

\vspace{1em}

\noindent
\hangindent=2em
Battistoni G \textit{et al} 2016 The FLUKA code: an accurate simulation tool for particle therapy \textit{Front. Oncol.} \textbf{6} 116

\vspace{1em}

\noindent
\hangindent=2em
Bethe H A and Ashkin J 1953 Passage of radiations through matter \textit{Exp. Nucl. Phys.} \textbf{1} 166--251 (New York: Wiley)

\vspace{1em}

\noindent
\hangindent=2em
Bohr N 1940 Scattering and stopping of fission fragments \textit{Phys. Rev.} \textbf{58} 654

\vspace{1em}

\noindent
\hangindent=2em
Bortfeld T 1997 An analytical approximation of the Bragg curve for therapeutic proton beams \textit{Medical Physics} \textbf{24} 2024--2033

\vspace{1em}

\noindent
\hangindent=2em
Böhlen T T \textit{et al} 2014 The FLUKA code: developments and challenges for high energy and medical applications \textit{Nucl. Data Sheets} \textbf{120} 211--214

\vspace{1em}

\noindent
\hangindent=2em
Bragg W H and Kleeman R 1905 On the $\alpha$ particles of radium, and their loss of range in passing through various atoms and molecules \textit{Lond. Edinb. Dubl. Phil. Mag. J. Sci.} \textbf{10} 318--340

\vspace{1em}

\noindent
\hangindent=2em
Chu W T, Ludewigt B A and Renner T R 1993 Instrumentation for treatment of cancer using proton and light ion beams \textit{Rev. Sci. Instrum.} \textbf{64} 2055--2122

\vspace{1em}

\noindent
\hangindent=2em
Donahue W, Newhauser W D and Ziegler J F 2016 Analytical model for ion stopping power and range in the therapeutic energy interval for beams of hydrogen and heavier ions \textit{Phys. Med. Biol.} \textbf{61} 6570

\vspace{1em}

\noindent
\hangindent=2em
Dudouet J, Juliani D, Labalme M, Cussol D, Angélique J C, Braunn B, Colin J, Finck C, Fontbonne J M, Guérin H \textit{et al.} 2013 Double-differential fragmentation cross-section measurements of 95 MeV/nucleon $^{12}$C beams on thin targets for hadron therapy \textit{Physical Review C} \textbf{88} 024606 

\vspace{1em}

\noindent
\hangindent=2em
Durante M and Loeffler J S 2010 Charged particles in radiation oncology \textit{Nat. Rev. Clin. Oncol.} \textbf{7} 37--43

\vspace{0.10em}

\noindent
\hangindent=2em
Durante M and Paganetti H 2016 Nuclear physics in particle therapy: a review \textit{Rep. Prog. Phys.} \textbf{79} 096702

\vspace{0.10em}

\noindent
\hangindent=2em
Durante M, Orecchia R and Loeffler J S 2017 Charged-particle therapy in cancer: clinical uses and future perspectives \textit{Nat. Rev. Clin. Oncol.} \textbf{14} 483--495

\vspace{0.10em}

\noindent
\hangindent=2em
El Bekkouri H, Al Ibrahmi E M, El-Asery M, Bardane A, Sadoune Z, Chakir E M, Didi A 2023 A new study of Bragg curve of the ¹²C Ion at energies ranging 200–400 MeV/u with the contribution of secondary fragments in hadrontherapy using the PHITS Monte Carlo code \textit{International Conference on Advanced Intelligent Systems for Sustainable Development} 244–254 \vspace{1em}

\vspace{0.10em}

\noindent
\hangindent=2em
Evans R D 1985 \textit{The Atomic Nucleus}, Reprint Edition 1982 of 14th Printing 1972 (Malabar, Florida: Krieger)

\vspace{0.10em}

\noindent
\hangindent=2em
Faddegon B \textit{et al} 2020 The TOPAS tool for particle simulation, a Monte Carlo simulation tool for physics, biology and clinical research \textit{Phys. Med.} \textbf{72} 114--121

\vspace{1em}

\noindent
\hangindent=2em
Fossati P, Matsufuji N, Kamada T, Karger C P 2018 Radiobiological issues in prospective carbon ion therapy trials \textit{Medical Physics} \textbf{45} e1096–e1110 

\vspace{1em}

\noindent
\hangindent=2em
Fujitaka S, Fujii Y, Nihongi H, Nakayama S, Takashina M, Hamatani N, Tsubouchi T, Yagi M, Minami K, Ogawa K \textit{et al.} 2021 Physical and biological beam modeling for carbon beam scanning at Osaka Heavy Ion Therapy Center \textit{Journal of Applied Clinical Medical Physics} \textbf{22} 77–92 

\vspace{1em}

\noindent
\hangindent=2em
Furusawa Y, Fukutsu K, Aoki M, Itsukaichi H, Eguchi-Kasai K, Ohara H, Yatagai F, Kanai T, Ando K 2000 Inactivation of aerobic and hypoxic cells from three different cell lines by accelerated $^3$He-, $^{12}$C- and $^{20}$Ne-Ion beams \textit{Radiation Research} \textbf{154} 485--496

\vspace{1em}

\noindent
\hangindent=2em
Furuta T and Sato T 2021 Medical application of particle and heavy ion transport code system PHITS \textit{Radiol. Phys. Technol.} \textbf{14} 215--225

\vspace{1em}

\noindent
\hangindent=2em
Francis Z, Seif E, Incerti S, Champion C, Karamitros M, Bernal M A, Ivanchenko V N, Mantero A, Tran H N and El Bitar Z 2014 Carbon ion fragmentation effects on the nanometric level behind the Bragg peak depth \textit{Phys. Med. Biol.} \textbf{59} 7691

\vspace{1em}

\noindent
\hangindent=2em
Golovchenko A N, Skvarč J, Ilić R, Sihver L, Bamblevski V P, Tretyakova S P, Schardt D, Tripathi R K, Wilson J W and Bimbot R 1999 Fragmentation of 200 and 244 MeV/u carbon beams in thick tissue-like absorbers \textit{Nucl. Instrum. Methods Phys. Res. Sect. B} \textbf{159} 233--240

\vspace{1em}

\noindent
\hangindent=2em
Golovchenko A N, Skvarč J, Yasuda N, Giacomelli M, Tretyakova S P, Ilić R, Bimbot R, Toulemonde M and Murakami T 2002 Total charge-changing and partial cross-section measurements in the reactions of $\sim$ 110--250 MeV/nucleon $^{12}{C}$ in carbon, paraffin, and water \textit{Phys. Rev. C} \textbf{66} 014609

\vspace{1em}

\noindent
\hangindent=2em
Gradshteyn I S and Ryzhik I M 1980 \textit{Tables of Integrals, Series and Products}, corrected and enlarged edition (San Diego: Academic Press) p. 388

\vspace{1em}

\noindent
\hangindent=2em
Grün R, Friedrich T, Elsässer T, Krämer M, Zink K, Karger C P, Durante M, Engenhart-Cabillic R, Scholz M 2012 Impact of enhancements in the local effect model (LEM) on the predicted RBE-weighted target dose distribution in carbon ion therapy \textit{Phys. Med. Biol.} \textbf{57} 7261–7277 

\vspace{1em}

\noindent
\hangindent=2em
Gunzert-Marx K, Iwase H, Schardt D and Simon R S 2008 Secondary beam fragments produced by 200 MeV  $u^{-1}$ $^{12}{C}$ ions in water and their dose contributions in carbon ion radiotherapy \textit{New J. Phys.} \textbf{10} 075003

\vspace{1em}

\noindent
\hangindent=2em
Haettner E, Iwase H, Krämer M, Kraft G and Schardt D 2013 Experimental study of nuclear fragmentation of 200 and 400 MeV/u $^{12}{C}$ ions in water for applications in particle therapy \textit{Phys. Med. Biol.} \textbf{58} 8265

\vspace{1em}

\noindent
\hangindent=2em
Halıcılar F, Arık M and Erkol H 2024 A comparison of the acoustic waves generated in proton and carbon ion therapy \textit{Phys. Scr.} \textbf{99} 115302

\vspace{1em}

\noindent
\hangindent=2em
Hamad M K 2021 Bragg-curve simulation of carbon-ion beams for particle-therapy applications: a study with the GEANT4 toolkit \textit{Nucl. Eng. Technol.}

\vspace{1em}

\noindent
\hangindent=2em
Hüfner J, Schäfer K and Schürmann B 1975 Abrasion-ablation in reactions between relativistic heavy ions \textit{Phys. Rev. C} \textbf{12} 1888

\vspace{1em}

\noindent
\hangindent=2em
Hollmark M, Uhrdin J, Belkić D, Gudowska I, Brahme A 2004 Influence of multiple scattering and energy loss straggling on the absorbed dose distributions of therapeutic light ion beams: I. Analytical pencil beam model \textit{Phys. Med. Biol.} \textbf{49} 3247 

\vspace{1em}

\noindent
\hangindent=2em
Hollmark M, Gudowska I, Belki{\'c} D{\v{z}}, Brahme A and Sobolevsky N 2008 An analytical model for light ion pencil beam dose distributions: multiple scattering of primary and secondary ions \textit{Phys. Med. Biol.} \textbf{53} 3477

\vspace{1em}

\noindent
\hangindent=2em
Inaniwa T, Furukawa T, Kase Y, Matsufuji N, Toshito T, Matsumoto Y, Furusawa Y, Noda K 2010 Treatment planning for a scanned carbon beam with a modified microdosimetric kinetic model \textit{Phys. Med. Biol.}
\textbf{55} 6721–6737 

\vspace{1em}

\noindent
\hangindent=2em
Inaniwa T, Kanematsu N, Hara Y, Furukawa T, Fukahori M, Nakao M, Shirai T 2014 Implementation of a triple Gaussian beam model with subdivision and redefinition against density heterogeneities in treatment planning for scanned carbon-ion radiotherapy \textit{Phys. Med. Biol.} \textbf{59} 5361–5384 

\vspace{1em}

\noindent
\hangindent=2em
Inaniwa T, Kanematsu N 2015 A trichrome beam model for biological dose calculation in scanned carbon-ion radiotherapy treatment planning \textit{Phys. Med. Biol.} \textbf{60} 437–451 

\vspace{1em}

\noindent
\hangindent=2em
Inaniwa T, Kanematsu N, Hara Y, Furukawa T 2015a Nuclear-interaction correction of integrated depth dose in carbon-ion radiotherapy treatment planning \textit{Phys. Med. Biol.} \textbf{60} 421–435 

\vspace{1em}

\noindent
\hangindent=2em
Inaniwa T, Kanematsu N, Matsufuji N, Kanai T, Shirai T, Noda K, Tsuji H, Kamada T, Tsujii H 2015b Reformulation of a clinical-dose system for carbon-ion radiotherapy treatment planning at the National Institute of Radiological Sciences, Japan \textit{Phys. Med. Biol.}
\textbf{60} 3271–3286 

\vspace{1em}

\noindent
\hangindent=2em
J{\"a}kel O, Schulz-Ertner D, Karger C P, Nikoghosyan A and Debus J 2003 Heavy ion therapy: status and perspectives \textit{Technol. Cancer Res. Treat.} \textbf{2} 377--387

\vspace{0.10em}

\noindent
\hangindent=2em
Jones K C, Vander Stappen F, Sehgal C M and Avery S 2016 Acoustic time-of-flight for proton range verification in water \textit{Med. Phys.} \textbf{43} 5213--5224

\vspace{0.10em}

\noindent
\hangindent=2em
Kameoka S, Amako K, Iwai G, Murakami K, Sasaki T, Toshito T, Yamashita T, Aso T, Kimura A, Kanai T \textit{et al.} 2008 Dosimetric evaluation of nuclear interaction models in the Geant4 Monte Carlo simulation toolkit for carbon-ion radiotherapy \textit{Radiological Physics and Technology} \textbf{1} 183–187 

\vspace{1em}

\noindent
\hangindent=2em
Kanai T, Furusawa Y, Fukutsu K, Itsukaichi H, Eguchi-Kasai K, Ohara H 1997 Irradiation of mixed beam and design of spread-out Bragg peak for heavy-ion radiotherapy \textit{Radiation Research} \textbf{147} 78–85

\vspace{1em}

\noindent
\hangindent=2em
Kanai T, Endo M, Minohara S, Miyahara N, Koyama-Ito H, Tomura H, Matsufuji N, Futami Y, Fukumura A, Hiraoka T and others 1999 Biophysical characteristics of HIMAC clinical irradiation system for heavy-ion radiation therapy \textit{Int. J. Radiat. Oncol. Biol. Phys.} \textbf{44} 201--210

\vspace{0.10em}

\noindent
\hangindent=2em
Kamada T, Tsujii H, Blakely E A, Debus J, De Neve W, Durante M, Jäkel O, Mayer R, Orecchia R, Pötter R and others 2015 Carbon ion radiotherapy in Japan: an assessment of 20 years of clinical experience \textit{Lancet Oncol.} \textbf{16} e93--e100

\vspace{0.10em}

\noindent
\hangindent=2em
Karger C P, Peschke P 2018 RBE and related modeling in carbon-ion therapy \textit{Phys. Med. Biol.} \textbf{63} 01TR02 

\vspace{1em}

\noindent
\hangindent=2em
Kase Y, Kanai T, Matsumoto Y, Furusawa Y, Okamoto H, Asaba T, Sakama M, Shinoda H 2006 Microdosimetric measurements and estimation of human cell survival for heavy-ion beams \textit{Radiation Research} \textbf{166} 629–638 

\vspace{1em}

\noindent
\hangindent=2em
Kelleter L and Jolly S 2020 A mathematical expression for depth-light curves of therapeutic proton beams in a quenching scintillator \textit{Medical Physics} \textbf{47} 2300--2308

\vspace{1em}

\noindent
\hangindent=2em
Kempe J and Brahme A 2010 Analytical theory for the fluence, planar fluence, energy fluence, planar energy fluence and absorbed dose of primary particles and their fragments in broad therapeutic light ion beams \textit{Physica Medica} \textbf{26} 6--16

\vspace{1em}

\noindent
\hangindent=2em
Kim J, Park J M and Wu H-G 2020 Carbon ion therapy: a review of an advanced technology \textit{Prog. Med. Phys.} \textbf{31} 71--80

\vspace{1em}

\noindent
\hangindent=2em
Kozłowska W S \textit{et al} 2019 FLUKA particle therapy tool for Monte Carlo independent calculation of scanned proton and carbon ion beam therapy \textit{Phys. Med. Biol.} \textbf{64} 075012

\vspace{1em}

\noindent
\hangindent=2em
Kraft G, Scholz M, Bechthold U 1999 Tumor therapy and track structure \textit{Radiation and Environmental Biophysics} \textbf{38} 229–237 

\vspace{1em}

\noindent
\hangindent=2em
Kraft G 2000 Tumor therapy with heavy charged particles \textit{Prog. Part. Nucl. Phys.} \textbf{45} S473--S544

\vspace{1em}

\noindent
\hangindent=2em
Kr{\"a}mer M \textit{et al} 2000 Treatment planning for heavy-ion radiotherapy: physical beam model and dose optimization \textit{Phys. Med. Biol.} \textbf{45} 3299

\vspace{1em}

\noindent
\hangindent=2em
Kr{\"a}mer M and Scholz M 2000 Treatment planning for heavy-ion radiotherapy: Calculation and optimization of biologically effective dose \textit{Phys. Med. Biol.} \textbf{45} 3319

\vspace{1em}

\noindent
\hangindent=2em
Kundrat P 2007 A semi-analytical radiobiological model may assist treatment planning in light ion radiotherapy \textit{Phys. Med. Biol.} \textbf{52} 6813

\vspace{1em}

\noindent
\hangindent=2em
Liu H \textit{et al} 2017 A preliminary Monte Carlo study for the treatment head of a carbon-ion radiotherapy facility using TOPAS \textit{EPJ Web Conf.} \textbf{153} 04018

\vspace{1em}

\noindent
\hangindent=2em
Loeffler J S, Durante M 2013 Charged particle therapy—optimization, challenges and future directions \textit{Nature Reviews Clinical Oncology} \textbf{10} 411–424 

\vspace{1em}

\noindent
\hangindent=2em
Lysakovski P, Kopp B, Tessonnier T, Mein S, Ferrari A, Haberer T, Debus J, Mairani A 2024 Development and validation of MonteRay, a fast Monte Carlo dose engine for carbon ion beam radiotherapy \textit{Medical Physics} \textbf{51} 1433–1449 

\vspace{1em}

\noindent
\hangindent=2em
Malouff T D, Mahajan A, Krishnan S, Beltran C, Seneviratne D S and Trifiletti D M 2020 Carbon ion therapy: a modern review of an emerging technology \textit{Front. Oncol.} \textbf{10} 82

\vspace{1em}

\noindent
\hangindent=2em
Matsufuji N, Fukumura A, Komori M, Kanai T and Kohno T 2003 Influence of fragment reaction of relativistic heavy charged particles on heavy-ion radiotherapy \textit{Phys. Med. Biol.} \textbf{48} 1605

\vspace{1em}

\noindent
\hangindent=2em
Mohamad O, Sishc B J, Saha J, Pompos A, Rahimi A, Story M D, Davis A J and Kim D W N 2017 Carbon ion radiotherapy: a review of clinical experiences and preclinical research, with an emphasis on DNA damage/repair \textit{Cancers} \textbf{9} 66

\vspace{0.10em}

\noindent
\hangindent=2em
Nakaji T, Kanai T, Takashina M, Matsumura A, Osaki K, Yagi M, Tsubouchi T, Hamatani N, Ogawa K 2022 Clinical dose assessment for scanned carbon-ion radiotherapy using linear energy transfer measurements and Monte Carlo simulations \textit{Phys. Med. Biol.} \textbf{67} 245021 

\vspace{1em}

\noindent
\hangindent=2em
Nandy M 2021 Secondary radiation in ion therapy and theranostics: a review \textit{Frontiers in Physics} \textbf{8} 598257

\vspace{1em}

\noindent
\hangindent=2em
Newhauser W D and Zhang R 2015 The physics of proton therapy \textit{Phys. Med. Biol.} \textbf{60} R155

\vspace{0.10em}

\noindent
\hangindent=2em
Nichelatti E, Ronsivalle C, Piccinini M, Picardi L and Montereali R M 2019 An analytical approximation of proton Bragg curves in lithium fluoride for beam energy distribution analysis \textit{Nucl. Instrum. Methods Phys. Res. B} \textbf{446} 29--36

\vspace{1em}

\noindent
\hangindent=2em
Niita K \textit{et al} 2010 PHITS: Particle and heavy ion transport code system, version 2.23 \textit{J. Nucl. Sci. Technol.}

\vspace{1em}

\noindent
\hangindent=2em
Parodi K, Mairani A, Brons S, Hasch B G, Sommerer F, Naumann J, Jäkel O, Haberer T, Debus J 2012 Monte Carlo simulations to support start-up and treatment planning of scanned proton and carbon ion therapy at a synchrotron-based facility \textit{Phys. Med. Biol.} \textbf{57} 3759–3774 

\vspace{1em}

\noindent
\hangindent=2em
Parodi K, Mairani A, Sommerer F 2013 Monte Carlo-based parametrization of the lateral dose spread for clinical treatment planning of scanned proton and carbon ion beams \textit{Journal of Radiation Research} \textbf{54} i91–i96 

\vspace{1em}

\noindent
\hangindent=2em
Perl J \textit{et al} 2012 TOPAS: an innovative proton Monte Carlo platform for research and clinical applications \textit{Med. Phys.} \textbf{39} 6818--6837

\vspace{1em}

\noindent
\hangindent=2em
Pshenichnov I, Mishustin I, Greiner W 2008 Comparative Study of Depth–Dose Distributions for Beams of Light and Heavy Nuclei in Tissue-Like Media \textit{Nuclear Instruments and Methods in Physics Research Section B: Beam Interactions with Materials and Atoms} \textbf{266} 1094–1098

\vspace{1em}

\noindent
\hangindent=2em
Pshenichnov I, Botvina A, Mishustin I and Greiner W 2010 Nuclear fragmentation reactions in extended media studied with Geant4 toolkit \textit{Nucl. Instrum. Methods Phys. Res. B} \textbf{268} 604--615

\vspace{1em}

\noindent
\hangindent=2em
Robert C \textit{et al} 2013 Distributions of secondary particles in proton and carbon-ion therapy: a comparison between GATE/Geant4 and FLUKA Monte Carlo codes \textit{Phys. Med. Biol.} \textbf{58} 2879

\vspace{1em}

\noindent
\hangindent=2em
Sato T \textit{et al} 2018 Features of particle and heavy ion transport code system (PHITS) version 3.02 \textit{J. Nucl. Sci. Technol.} \textbf{55} 684--690

\vspace{1em}

\noindent
\hangindent=2em
Schall I, Schardt D, Geissel H, Irnich H, Kankeleit E, Kraft G, Magel A, Mohar M F, Münzenberg G and Nickel F 1996 Charge-changing nuclear reactions of relativistic light-ion beams ($5 \leq Z \leq 10$) passing through thick absorbers \textit{Nucl. Instrum. Methods Phys. Res. Sect. B} \textbf{117} 221--234

\vspace{1em}

\noindent
\hangindent=2em
Schardt D, Elsässer T and Schulz-Ertner D 2010 Heavy-ion tumor therapy: physical and radiobiological benefits \textit{Rev. Mod. Phys.} \textbf{82} 383--425

\vspace{0.10em}

\noindent
\hangindent=2em
Schulz-Ertner D, Jäkel O, Schlegel W 2006 Radiation therapy with charged particles \textit{Seminars in Radiation Oncology} \textbf{16} 249–259 

\vspace{1em}

\noindent
\hangindent=2em
Schulz-Ertner D, Tsujii H 2007 Particle Radiation Therapy Using Proton and Heavier Ion Beams \textit{Journal of Clinical Oncology} \textbf{25} 953–964 

\vspace{1em}

\noindent
\hangindent=2em
Scholz M, Kraft G 1996 Track structure and the calculation of biological effects of heavy charged particles \textit{Advances in Space Research} \textbf{18} 5–14 

\vspace{1em}

\noindent
\hangindent=2em
Scholz M, Kellerer A M, Kraft-Weyrather W, Kraft G 1997 Computation of cell survival in heavy ion beams for therapy: the model and its approximation \textit{Radiation and Environmental Biophysics} \textbf{36} 59–66 

\vspace{1em}

\noindent
\hangindent=2em
Serber R 1947 Nuclear reactions at high energies \textit{Phys. Rev.} \textbf{72} 1114

\vspace{1em}

\noindent
\hangindent=2em
Sigmund P, Schinner A and Paul H 2009 Errata and addenda for ICRU Report 73, Stopping of ions heavier than helium \textit{J. ICRU} \textbf{5} 1--10

\vspace{1em}

\noindent
\hangindent=2em
Sigmund P 2014 \textit{Particle Penetration and Radiation Effects}, Vol. 2, \textit{Springer Series in Solid-State Sciences} \textbf{179} (Berlin: Springer)

\vspace{1em}

\noindent
\hangindent=2em
Suzuki M, Kase Y, Yamaguchi H, Kanai T, Ando K 2000 Relative biological effectiveness for cell-killing effect on various human cell lines irradiated with heavy-ion medical accelerator in Chiba (HIMAC) carbon-ion beams \textit{International Journal of Radiation Oncology Biology Physics} \textbf{48} 241–250

\vspace{1em}

\noindent
\hangindent=2em
Tinganelli W and Durante M 2020 Carbon ion radiobiology \textit{Cancers} \textbf{12} 3022

\vspace{0.10em}

\noindent
\hangindent=2em
Tsuji H, Yanagi T, Ishikawa H, Kamada T, Mizoe J, Kanai T, Morita S, Tsujii H \textit{et al.} 2005 Hypofractionated radiotherapy with carbon ion beams for prostate cancer \textit{International Journal of Radiation OncologyBiologyPhysics} \textbf{63} 1153–1160

\vspace{1em}

\noindent
\hangindent=2em
Tsujii H and Kamada T 2012 A review of update clinical results of carbon ion radiotherapy \textit{Jpn. J. Clin. Oncol.} \textbf{42} 670--685

\vspace{0.10em}

\noindent
\hangindent=2em
Vavilov P V 1957 Ionization losses of high-energy heavy particles \textit{Sov. Phys. JETP} \textbf{5}

\vspace{0.10em}

\noindent
\hangindent=2em
Yang S, Zhao J, Zhuo W, Shen H, Chen B 2021 Measurement of therapeutic $^{12}$C beam in a water phantom using CR-39 \textit{Journal of Radiological Protection} \textbf{41} 279–293 

\vspace{1em}

\noindent
\hangindent=2em
Yagi M, Tsubouchi T, Hamatani N, Takashina M, Maruo H, Fujitaka S, Nihongi H, Ogawa K, Kanai T 2022 Commissioning a newly developed treatment planning system, VQA Plan, for fast-raster scanning of carbon-ion beams \textit{PLOS ONE} \textbf{17} e0268087 

\vspace{1em}

\noindent
\hangindent=2em
Yagi M, Tsubouchi T, Hamatani N, Takashina M, Saruwatari N, Minami K, Wakisaka Y, Fujitaka S, Hirayama S, Nihongi H \textit{et al.} 2023 Validation of robust radiobiological optimization algorithms based on the mixed beam model for intensity-modulated carbon-ion therapy \textit{PLOS ONE} \textbf{18} e0288545 

\vspace{1em}

\noindent
\hangindent=2em
Ying C K, Bolst D, Rosenfeld A and Guatelli S 2019 Characterization of the mixed radiation field produced by carbon and oxygen ion beams of therapeutic energy: a Monte Carlo simulation study \textit{J. Med. Phys.} \textbf{44} 263--269

\vspace{1em}

\noindent
\hangindent=2em
Ying C K, Bolst D, Tran L T, Guatelli S, Rosenfeld A B and Kamil W A 2017 Contributions of secondary fragmentation by carbon ion beams in water phantom: Monte Carlo simulation \textit{J. Phys.: Conf. Ser.} \textbf{851} 012033

\vspace{1em}

\noindent
\hangindent=2em
Yogo K, Tsuneda M, Horita R, Souda H, Matsumura A, Ishiyama H, Hayakawa K, Kanai T, Yamamoto S 2021 Three-dimensional dose-distribution measurement of therapeutic carbon-ion beams using a ZnS scintillator sheet \textit{Journal of Radiation Research} \textbf{62} 825–832 

\vspace{1em}

\noindent
\hangindent=2em
Yoon E, Kim J, Park J M, Choi C H, Jung S 2023 Extension of matRad with a modified microdosimetric kinetic model for carbon ion treatment planning: Comparison with Monte Carlo calculation \textit{Medical Physics} \textbf{50} 5884–5896

\vspace{1em}

\noindent
\hangindent=2em
Zhang X \textit{et al} 2011 Parameterization of multiple Bragg curves for scanning proton beams using simultaneous fitting of multiple curves \textit{Phys. Med. Biol.} \textbf{56} 7725

\vspace{1em}

\noindent
\hangindent=2em
Zhang H, Dai Z, Liu X, Chen W, Ma Y, He P, Dai T, Shen G, Yuan P, Li Q 2018 A novel pencil beam model for carbon-ion dose calculation derived from Monte Carlo simulations \textit{Physica Medica} \textbf{55} 15–24

\vspace{1em}

\noindent
\hangindent=2em
Zhang J, Liang Y and Yang C 2022a A primary proton integral depth dose calculation model corrected with straight scattering track approximation \textit{Radiat. Phys. Chem.} \textbf{201} 110283

\vspace{1em}

\noindent
\hangindent=2em
Zhang H, Li Q, Liu X, Ma Y, He P, Shen G, Li Z, Chen W, Niu R, Dai Z \textit{et al.} 2022b Validation and testing of a novel pencil-beam model derived from Monte Carlo simulations in carbon-ion treatment planning for different scenarios \textit{Physica Medica} \textbf{99} 1–9 

\vspace{1em}

\noindent
\hangindent=2em
Zhang J, Liang Y and Yang C 2023 Proton beam secondary depth-dose calculation with a secondary propagation model \textit{Radiat. Phys. Chem.} \textbf{204} 110679

\vspace{1em}

\noindent
\hangindent=2em
Ziegler J F, Ziegler M D and Biersack J P 2010 SRIM—The stopping and range of ions in matter (2010) \textit{Nucl. Instrum. Methods Phys. Res. B} \textbf{268} 1818--1823

\vspace{1em}

\end{document}